\documentclass[12pt,english]{article}
\usepackage[T1]{fontenc}
\usepackage[latin9]{inputenc}
\usepackage{xcolor}
\usepackage{mathtools}
\usepackage{amsmath}
\usepackage{amssymb}
\usepackage{stmaryrd}
\usepackage{graphicx}

\makeatletter
\usepackage{jheppub}
\def\@fpheader{\relax}
\allowdisplaybreaks[1]

\makeatother

\usepackage{babel}
\begin{document}


\title{A generalized uncertainty-inspired quantum black hole}

\author[a]{Federica Fragomeno,}
\author[a,b]{Douglas M. Gingrich,}
\author[c,d]{Samantha Hergott,}
\author[a,e,f]{Saeed~Rastgoo,}
\author[a]{Evan Vienneau,}
\affiliation[a]{Department of Physics, University of Alberta, Edmonton, Alberta T6G 2G1, Canada} 
\affiliation[b]{TRIUMF, Vancouver, BC V6T 2A3, Canada}
\affiliation[c]{Department of Physics and Astronomy, York University, Toronto, Ontario M3J 1P3, Canada}
\affiliation[d]{Perimeter Institute for Theoretical Physics, Waterloo, Ontario N2L 2Y5, Canada}
\affiliation[e]{Department of Mathematical and Statistical Sciences, University of Alberta, Edmonton, Alberta T6G 2G1, Canada}
\affiliation[f]{Theoretical Physics Institute, University of Alberta, Edmonton, Alberta T6G 2G1, Canada}
\emailAdd{ffragome@ualberta.ca}
\emailAdd{gingrich@ualberta.ca}
\emailAdd{sherrgs@yorku.ca}
\emailAdd{srastgoo@ualberta.ca}
\emailAdd{eviennea@ualberta.ca}

\abstract{

We derive the full spacetime metric of a generalized uncertainty-inspired
quantum black hole. We examine a previous model of the interior in this approach and show that extending its metric to the full
spacetime leads to serious issues in the asymptotic region. To remedy
this, we introduce an ``improved scheme'' mimicking a similar prescription
used in loop quantum gravity, where the quantum parameters are made momentum-dependent. Under this scheme, we rework the interior
of the black hole and extend it to the full spacetime. We find that
the resulting metric is asymptotically flat and its associated Kretschmann
scalar is regular everywhere. We also show that the null expansion and Raychaudhuri
equation are regular everywhere in this spacetime, implying that
the classical singularity is resolved.

}
\maketitle

\section{Introduction}

Black holes are considered to be one of the places where new physics
can emerge. Thus studying quantum black holes is one of the most
important tasks in quantum gravity, in the hope of obtaining
hints about the nature of quantum spacetime. This
is even more crucial in the age in which there are hopes that several
near-future experiments could observe quantum gravity
signatures in astrophysical phenomena, particularly from black holes
\cite{Addazi:2021xuf,LISA:2022kgy,LISACosmologyWorkingGroup:2022jok,AlvesBatista:2023wqm}.

Quantum black holes have been studied in many approaches. In a non-perturbative
approach to quantum gravity, called loop quantum gravity (LQG)~\cite{Thiemann:2007pyv},
various models have been put forward. These models are based on a
classical phase space with the Ashtekar-Barbero (AB) connection as the  configuration
variable, and its conjugate momentum the densitized triad. The latter
can be thought of as components of the spatial metric. One would then
transition to loop classical theory by introducing holonomies of the
AB connection as the configuration variable and the flux of densitized
triads over 2-surfaces as their momenta. One then quantizes these
variables over a suitable Hilbert space. The black hole models in LQG usually follow this process
by first symmetry-reducing the classical theory. They include
models \cite{Ashtekar:2005qt,Bohmer:2007wi,Chiou:2008nm,Morales-Tecotl:2018ugi,Blanchette:2020kkk}
that treat the interior as a Kantowski-Sachs cosmology, the full spacetime \cite{Kelly:2020uwj,Gambini:2020nsf,Ashtekar:2018lag,Bodendorfer:2019nvy,Modesto:2008im,Bojowald:2020dkb},
or further symmetry-reduced 2D systems \cite{Gambini:2009vp,Corichi:2015vsa,Corichi:2016nkp}.
These models either holonomize or regularize the system as described
above using LQG techniques, or use polymer quantization techniques \cite{Ashtekar:2002sn,Morales-Tecotl:2016ijb,Tecotl:2015cya}.
Some of the models follow the procedure without actually quantizing
the holonomies and fluxes \cite{Kelly:2020uwj,Ashtekar:2018lag,Bodendorfer:2019nvy,Modesto:2008im,Bojowald:2020dkb},
obtaining an ``effective'' metric just from regularization. Others proceed to  quantization and extract
the effective metric by obtaining the expectation values of the metric components
in certain states that are peaked around classical solutions \cite{Gambini:2020nsf}. 

Another approach to quantum gravity is the so called generalized  uncertainty principle (GUP) approach, sometimes also known as minimal uncertainty approach \cite{Kempf:1994su,Bosso:2023aht}.
This approach is based on the observation that the product of uncertainties
in configuration and momenta have a certain relation to the algebra
of these variables. Hence, by demanding certain minimal uncertainty
in either the configuration or the momenta, the algebra will be modified. This approach has been mostly studied in the case of finite
number of degrees of freedom rather than in  field theories.
Hence, directly applying it to the full spacetime of a black hole
is still not well understood. However, this model can be used, and
has been used, to study the interior of the Schwarzschild black hole
in AB variables, both in the effective regime \cite{Bosso:2020ztk,Blanchette:2021vid}
including a comparison with LQG \cite{Rastgoo:2022mks}, and in the quantum
regime \cite{Bosso:2023fnb}. 

There have also been several works studying some aspects of black holes in GUP, particularly their thermodynamics \cite{Lambiase:2017adh, Scardigli:2014qka, Casadio:2016fev, Jizba:2009qf, Carr:2015nqa, Ong:2018zqn, Buoninfante:2019fwr}, and some of them have put forward certain forms of GUP-deformed  components of a Schwrzschild-like metric up to the first order in GUP parameters, albeit using heuristic methods \cite{Lambiase:2017adh,Scardigli:2014qka, Carr:2015nqa}. However, to our knowledge, there have been no systematic derivations of a full GUP metric (i.e., not perturbatively up to a certain order in GUP parameters, and not solely based on heuristic arguments) until now. This is where our work fills the gap. 

In this work, we derive  the full spacetime
metric of a GUP-modified Schwarzschild black hole
using a combination of techniques from both GUP and LQG. To avoid dealing with the full spacetime directly, which is a field
theory, we start by treating the interior written in AB
variables using a GUP-modified algebra. Then we
extend the resulting metric of the interior to
the full spacetime. This is similar to the approach used in some of the proposed models in LQG \cite{Ashtekar:2018lag,Bodendorfer:2019nvy,Modesto:2005zm}. However,
the resulting metric suffers from some of  the asymptotic issues encountered
in \cite{Ashtekar:2018lag} and discussed further in \cite{Ashtekar:2020ckv,Bouhmadi-Lopez:2019hpp}.
In order to remedy these issues, we follow the ``improved prescriptions''
in LQG by making the quantum parameters of the model,
which were originally constant, momentum-dependent. Remarkably this
prescription not only resolves the asymptotic issues but renders the
black hole regular.

The order of material in this work is as follows. In Sec.~\ref{sec:Classical-Schwarzschild-interior}
we present a brief overview of the dynamics of the interior of the
Schwarzschild black hole in AB variables. In Sec.~\ref{sec:Non-improved-metric}
we show how the GUP treatment of the interior without
considering momentum-dependent quantum parameters, leads to a full
spacetime metric and a Kretschmann scalar with asymptotic issues. Sec.~\ref{sec:Iimproved-metric} is dedicated to reworking the interior
with the improved GUP approach (where the modifications to the algebra are made momentum dependent to obtain an improved version in which the classical and asymptotic limits are correct), extending it
to the full spacetime, and showing that not only this metric has all the desired
asymptotic properties, but also its  Kretschmann scalar
is regular everywhere. We also work out some of the properties of this metric including the effective
stress-energy tensor, obtained if we would assume Einstein's equations are
valid. In Sec.~\ref{sec:Geodesics} we study several properties of
geodesics in this spacetime, including the photon spheres, the velocity of the infalling observer, and more importantly, the expansion scalar and the Raychaudhuri equation. We argue that all these results point to the resolution of the classical singularity which is demonstrated by showing that the expansion scalar and the Raychaudhuri equation are
regular everywhere. Finally, in Sec.~\ref{sec:Conclusion}, we summarize
our results and make some concluding remarks.

\section{Classical Schwarzschild interior in Ashtekar-Barbero variables\label{sec:Classical-Schwarzschild-interior}}

We begin by reviewing the interior solution for a Schwarzschild black hole in Ashtekar-Barbero variables. 
The interior of a static spherically symmetric black hole in Ashtekar-Barberos
variables in Schwarzschild coordinates can be described by the metric
\cite{Ashtekar:2005qt}
\begin{equation}
d\tilde{s}^{2}=-\tilde{N}(\tilde{T})^{2}d\tilde{T}^{2}+\frac{\tilde{p}_{b}^{2}(\tilde{T})}{L_{0}^{2}\left|\tilde{p}_{c}(\tilde{T})\right|}d\tilde{r}^{2}+\left|\tilde{p}_{c}(\tilde{T})\right|\left(d\tilde{\theta}^{2}+\sin^{2}(\tilde{\theta})d\tilde{\phi}^{2}\right),\label{eq:gen-metric-in-T}
\end{equation}
where tilde symbols refer to the interior coordinates and variables, and
$\tilde{T}$ is timelike and $\tilde{r}$ is spacelike in the interior.
The function $\tilde{N}$ is the lapse function arising due to the
ADM (Arnowitt-Deser-Misner) decomposition of the interior spacetime. The timelike coordinate $\tilde{T}$
is a generic one that is associated with the choice of $\tilde{N}$,
and is generally not the timelike coordinate of the Schwarzschild
interior metric which we call $\tilde{t}$. The variables $\tilde{b},\,\tilde{c},\,\tilde{p}_{b}$
and $\tilde{p}_{c}$ are the components of the Ashtekar-Barbero connection
$A_{a}^{i}$ and the densitized triad $\mathcal{E}_{i}^{a}$, respectively
\cite{Ashtekar:2005qt}, adapted to the symmetry of the model:
\begin{eqnarray}
A_{a}^{i}\tau_{i}d\tilde{x}^{a} & = & \frac{\tilde{c}}{L_{0}}\tau_{3}d\tilde{r}+\tilde{b}\tau_{2}d\tilde{\theta}-\tilde{b}\tau_{1}\sin(\tilde{\theta})d\tilde{\phi}+\tau_{3}\cos(\tilde{\theta})d\tilde{\phi},\\
\mathcal{E}_{i}^{a}\tau_{i}\partial_{a} & = & \tilde{p}_{c}\tau_{3}\sin(\tilde{\theta})\partial_{\tilde{r}}+\frac{\tilde{p}_{b}}{L_{0}}\tau_{2}\sin(\tilde{\theta})\partial_{\tilde{\theta}}-\frac{\tilde{p}_{b}}{L_{0}}\tau_{1}\partial_{\tilde{\phi}}.
\end{eqnarray}
Hence, $\tilde{b}$ and $\tilde{c}$
are the configuration variables, and $\tilde{p}_{b}$ and $\tilde{p}_{c}$
are their associated conjugate momenta. Here $L_{0}$ is a fiducial
parameter introduced to make the symplectic structure, and hence the
definition of the Poisson brackets, well-defined. Clearly, no physical
observables should depend on $L_{0}$ or on its rescaling. The $\tau_{i}$
are the basis of the $su(2)$, and $i$ is an $SU(2)$ index, while
$a$ is a spatial index. The algebra of canonical variables, inherited
from the algebra of $A_{a}^{i}$ and $\mathcal{E}_{i}^{a}$, is
\begin{equation}
\{ \tilde{b},\tilde{p}_{b}\} = G\gamma \quad\textrm{and}\quad \left\{ \tilde{c},\tilde{p}_{c}\right\} = 2G\gamma,\label{eq:PB-classic}
\end{equation}
where $\gamma$ is the Barbero-Immirzi parameter \cite{Thiemann:2007pyv},
which is a free parameter in loop quantum gravity that is taken to
be real and positive, and can be fixed by, for example, computations of black
hole entropy. The classical Hamiltonian (constraint) under the relevant
symmetry reduction of this model becomes \cite{Ashtekar:2005qt}
\begin{equation}
\tilde{H}=-\frac{\tilde{N}\text{sgn}(\tilde{p}_{c})}{2G\gamma^{2}}\left[\left(\tilde{b}^{2}+\gamma^{2}\right)\frac{\tilde{p}_{b}}{\sqrt{\left|\tilde{p}_{c}\right|}}+2\tilde{b}\tilde{c}\sqrt{\left|\tilde{p}_{c}\right|}\text{sgn}(\tilde{p}_{c})\right].
\end{equation}
The classical Hamiltonian equations of motion $\dot{f}=\left\{ f,H\right\}$
for $\tilde{c}$ and $\tilde{p}_{c}$ can be decoupled from those of
$\tilde{b}$ and $\tilde{p}_{b}$ by choosing the lapse 
\begin{equation}
\tilde{N}(\tilde{T})=\frac{\gamma\,\text{sgn}(\tilde{p}_{c})\sqrt{|\tilde{p}_{c}(\tilde{T})|}}{\tilde{b}(\tilde{T})},\label{eq:lapse}
\end{equation}
which reduces the Hamiltonian to
\begin{equation}
\tilde{H}=-\frac{1}{2G\gamma}\left[\left(\tilde{b}^{2}+\gamma^{2}\right)\frac{\tilde{p}_{b}}{\tilde{b}}+2\tilde{c}\tilde{p}_{c}\right].\label{eq:H-tilde}
\end{equation}
Solving these equations of motion yield
\begin{align}
\tilde{b}(\tilde{T}) & =\pm\sqrt{e^{2C_{1}}e^{-\tilde{T}}-\gamma^{2}},\label{eq:EoM-b-T-not-fixed}\\
\tilde{p}_{b}(\tilde{T}) & =C_{2}e^{\frac{\tilde{T}}{2}}\sqrt{e^{2C_{1}}-\gamma^{2}e^{\tilde{T}}},\label{eq:EoM-pb-T-not-fixed}\\
\tilde{c}(\tilde{T}) & =\mp C_{3}e^{-2\tilde{T}},\label{eq:EoM-c-T-not-fixed}\\
\tilde{p}_{c}(\tilde{T}) & =\pm C_{4}e^{2\tilde{T}}.\label{eq:EoM-pc-T-not-fixed}
\end{align}
To fix the integration constants, we can compare the spatial components
of the metric (\ref{eq:gen-metric-in-T}) with the Schwarzschild interior
metric:
\begin{eqnarray}
\frac{\tilde{p}_{b}^{2}(\tilde{T})}{L_{0}^{2}\tilde{p}_{c}(\tilde{T})} & = & g_{rr}(\tilde{t})=\left(\frac{2GM}{t}-1\right),\label{eq:match-pb}\\
\tilde{p}_{c}(\tilde{T}) & = & g_{\theta\theta}(\tilde{t})=\frac{g_{\phi\phi}(\tilde{t})}{\sin^2(\tilde{\theta})}=\tilde{t}^{2}.\label{eq:match-pc}
\end{eqnarray}
We see that to match (\ref{eq:match-pc}) and (\ref{eq:EoM-pc-T-not-fixed}),
we need a coordinate transformation $\tilde{T}=\ln(\tilde{t})$
and a choice of $C_{4}=1$ (or $C_{4}=-1$ for the negative solution of
$\tilde{p}_{c}$). Notice that this means $\tilde{t}=0$ is not part of the manifold.  This also means that in Schwarzschild coordinates
$(\tilde{t},\tilde{r},\tilde{\theta},\tilde{\phi})$, the
general spherically symmetric metric corresponding to (\ref{eq:gen-metric-in-T})
is

\begin{equation}
d\tilde{s}^{2}=-\frac{\tilde{N}(\tilde{t})^{2}}{\tilde{t}^{2}}d\tilde{t}^{2}+\frac{\tilde{p}_{b}^{2}(\tilde{t})}{L_{0}^{2}\tilde{p}_{c}(\tilde{t})}d\tilde{r}^{2}+\tilde{p}_{c}(\tilde{t})\left(d\tilde{\theta}^{2}+\sin^{2}(\tilde{\theta})d\tilde{\phi}^{2}\right).\label{eq:gen-metric-in-t}
\end{equation}
The remainder of the integration constants in (\ref{eq:EoM-b-T-not-fixed})-(\ref{eq:EoM-c-T-not-fixed})
can be found by considering $\tilde{p}_{b}(\tilde{t} = 2GM)=0$
and by matching (\ref{eq:match-pb}) to the solutions -- all after the
coordinate transformation $\tilde{T}=\ln(\tilde{t})$.
Then the classical solutions in the Schwarzschild interior coordinates
become
\begin{align}
\tilde{b}(\tilde{t}) & =\gamma\sqrt{\frac{R_{s}}{\tilde{t}}-1},\label{eq:class-sol-b-int-t}\\
\tilde{p}_{b}(\tilde{t}) & =L_{0}\tilde{t}\sqrt{\frac{R_{s}}{\tilde{t}}-1},\\
\tilde{c}(\tilde{t}) & =-\frac{\gamma L_{0}R_{s}}{2\tilde{t}^{2}},\\
\tilde{p}_{c}(\tilde{t}) & =\tilde{t}^{2},\label{eq:class-sol-pc-int-t}
\end{align}
where $R_{s}=2GM$ is the Schwarzschild radius.

In the classical regime, the Kretschmann scalar in variables \eqref{eq:class-sol-b-int-t}-\eqref{eq:class-sol-pc-int-t}
becomes 
\begin{equation}
K_{\text{class}}=\frac{12\left(\tilde{b}^{2}+\gamma^{2}\right)^{2}}{\gamma^{4}\tilde{p}_{c}^{2}}.\label{K-Classic}
\end{equation}
Notice that $|\tilde{p}_{c}|$ is the square of the radius of 2-spheres in these coordinates. The Kretschmann scalar above clearly diverges for $\tilde{p}_{c}\to0$, which is the position
where the classical singularity resides. 

\section{Non-improved effective metric \label{sec:Non-improved-metric}}

\subsection{Non-improved canonical variables and dynamics of the interior }

We are now going to implement modifications to the Poisson algebra
according to the GUP approach~\cite{Kempf:1994su, Bosso:2023aht} by choosing a quadratic modification in configuration variables as 
\begin{align}
\left\{ b,p_{b}\right\} & = G\gamma\left(1+\beta_{b}b^{2}\right),\label{eq:deform-1}\\
\left\{ c,p_{c}\right\} & = 2 G\gamma\left(1+\beta_{c}c^{2}\right),\label{eq:deform-2}
\end{align}
where $\beta_b$ and $\beta_c$ are small dimensionless parameters, also known as GUP parameters.

The GUP approach is based on the
observation that the uncertainty in the configuration
variable $q$ and its conjugate momentum $p$, is related to their
algebra via
\begin{equation}
\Delta q\Delta p\geq\frac{\hbar}{2}|\langle[\hat{q},\hat{p}]\rangle|.
\end{equation}
Therefore a modification to the algebra such as
\begin{equation}
[\hat{q},\hat{p}]=i\hbar\left(1+\beta\left\langle q^{2}\right\rangle +\alpha\left\langle p^{2}\right\rangle \right)\label{eq:qpcommut}
\end{equation}
leads to a modified uncertainty relation \cite{Kempf:1994su, Bosso:2023aht}
\begin{equation}
\Delta q\Delta p\geq\frac{\hbar}{2}\left[1+\beta(\Delta q)^{2}+\alpha(\Delta p)^{2}+\beta\left\langle q\right\rangle ^{2}+\alpha\left\langle p\right\rangle ^{2}\right].
\end{equation}
This in turn implies the existence of a minimal uncertainty in both
$q$ and $p$, given that $\alpha>0$ and $\beta>0$ \cite{Kempf:1994su, Bosso:2023aht}. Applied to the gravitational variables as we have done above, it leads to a minimal uncertainty in the momenta which are the components of the triad, and ultimately the  metric. So given that $\beta>0$, this modification of algebra leads to some sort of minimum value associated to the metric.

This approach however, will encounter several issues if it
is only restricted to positive deformation parameters $\alpha$ and
$\beta$. Some of the well-know issues are the followings. Positive
deformation parameters result in the Chandrasekhar limit no longer existing,
thus making arbitrarily large white dwarfs allowed \cite{Rashidi:2015rro,Mathew:2020wnx,Ong:2018zqn}.
The computation of temperature of the Hawking radiation could lead
to negative deformation parameters \cite{Lambiase:2017adh,Scardigli:2014qka,Buoninfante:2019fwr}.
Furthermore, in a systematic study of the interior dynamics, the only
way to remove the singularity seems to enforce the negative sign on the deformation
parameters \cite{Bosso:2020ztk,Blanchette:2021vid,Rastgoo:2022mks}.
Finally an analysis based on horizon quantum mechanics also leads to
the fact that these GUP parameters should be negative \cite{Casadio:2016fev}.

If one extends this method, particularly
\eqref{eq:qpcommut}, to negative values of GUP parameters, one
will not obtain a minimal uncertainty anymore, but the aforementioned issues will
be resolved. The negativity of deformation parameters hints that the physical space-time has
actually a lattice or granular structure at the Planck
scale, essentially a crystal-like universe whose lattice spacing is
of the order of Planck length \cite{Jizba:2009qf}. We will discuss the particular case of our model in more detail at the end of Sec. \ref{Sec:transit-to-improved-metric}.

Although \eqref{eq:deform-1} and \eqref{eq:deform-2} do not modify the Hamiltonian, they lead to effective equations of motion that are different
from the classical ones due to modification to the Poisson algebra (see appendix \ref{app:Interior-EoM}). The solution to these modified equations
of motion are \cite{Bosso:2020ztk,Blanchette:2021vid,Rastgoo:2022mks}
\begin{align}
\tilde{b}= & \frac{\gamma\sqrt{R_{s}t^{\beta_{b}\gamma^{2}}-t\left(\gamma^{2}R_{s}\right)^{\beta_{b}\gamma^{2}}}}{\sqrt{t\left(\gamma^{2}R_{s}\right)^{\beta_{b}\gamma^{2}}-\beta_{b}\gamma^{2}R_{s}t^{\beta_{b}\gamma^{2}}}}\label{eq:GUP-int-b}\, ,\\
\tilde{p}_{b}=~ & \gamma L_{0}t^{-\beta_{b}\gamma^{2}}\sqrt{R_{s}t^{\beta_{b}\gamma^{2}}-t\left(\gamma^{2}R_{s}\right)^{\beta_{b}\gamma^{2}}}\sqrt{t\left(\gamma^{2}R_{s}\right)^{\beta_{b}\gamma^{2}}-\beta_{b}\gamma^{2}R_{s}t^{\beta_{b}\gamma^{2}}}\, ,\\
\tilde{c}= & -\frac{\gamma L_{0}R_{s}}{2\sqrt{t^{4}-\frac{1}{4}\beta_{c}\gamma^{2}L_{0}^{2}R_{s}^{2}}}\, ,\\
\tilde{p}_{c}=~ & \sqrt{t^{4}-\frac{1}{4}\beta_{c}\gamma^{2}L_{0}^{2}R_{s}^{2}}\label{eq:GUP-int-pc}\, .
\end{align}
The constants of integration in these solutions have been fixed by
matching the classical limit of these solutions when $\beta_{b}\to0$ and $\beta_{c}\to0$,
with the actual classical solutions (\ref{eq:class-sol-b-int-t})-(\ref{eq:class-sol-pc-int-t}).
Replacing the solutions into the metric (\ref{eq:gen-metric-in-t}) together
with (\ref{eq:lapse}) (written in $\tilde{t}$) yields the effective
interior metric.

\subsection{Full spacetime extension of the non-improved metric}

Since the extended (interior and exterior) metric of the classical Schwarzschild spacetime can
be derived by switching the timelike and spacelike coordinates of
the Schwrazschild interior, as a first attempt, we try to apply the
same concept to the interior metric in the GUP approach
to obtain the full spacetime metric. In other words, we switch $\tilde{t}\to r$
and $\tilde{r}\to t$ (and for consistency in notation $\tilde{\theta}\to\theta$
and $\tilde{\phi}\to\phi$) in the solutions (\ref{eq:GUP-int-b})-(\ref{eq:GUP-int-pc})
and in the metric (\ref{eq:gen-metric-in-t}). Note that since we mentioned before that $\tilde{t}=0$ is not part of the manifold, the aforementioned switch means now $r=0$ is not part of the manifold. The extended (interior
and exterior) spacetime metric is now
\begin{equation}
ds^{2}=\frac{\check{p}_{b}^{2}\left(r\right)}{L_{0}^{2}\check{p}_{c}\left(r\right)}dt^{2}-\frac{1}{r^{2}}\frac{\gamma^{2}\check{p}_{c}\left(r\right)}{\check{b}^{2}\left(r\right)}dr^{2}+\check{p}_{c}\left(r\right)\left(d\theta^{2}+\sin^{2}\left(\theta\right)d\phi^{2}\right),\label{eq:metric-check}
\end{equation}
where we have denoted the extended canonical variables by check symbols.
From now on we will consider both interior and exterior as described
by coordinates $(t,\,r,\,\theta,\,\phi)$, and keep in mind that $t$ and $r$
are timelike and spacelike in the exterior, respectively, and spacelike
and timelike in the interior, respectively. 

In order to write the canonical variables (\ref{eq:GUP-int-b})-(\ref{eq:GUP-int-pc})
and the full spacetime metric in a concise way, we define the following
quantities
\begin{align}
\check{Q}_{b}= & \left|\beta_{b}\right|\gamma^{2},\\
\check{Q}_{c}= & \left|\beta_{c}\right|\gamma^{2},
\end{align}
and
\begin{eqnarray}
\check{\lambda}\left(r\right) & = & r\left(\gamma^{2}R_{s}\right)^{\text{sgn}\left(\beta_{b}\right)\check{Q}_{b}}-R_{s}r^{\text{sgn}\left(\beta_{b}\right)\check{Q}_{b}},\\
\check{\mu}\left(r\right) & = & r\left(\gamma^{2}R_{s}\right)^{\text{sgn}\left(\beta_{b}\right)\check{Q}_{b}}-\text{sgn}\left(\beta_{b}\right)\check{Q}_{b}R_{s}r^{\text{sgn}\left(\beta_{b}\right)\bar{Q}_{b}},\\
\check{\xi}\left(r\right) & = & 4r^{4}-L_{0}^{2}\,\text{sgn}\left(\beta_{c}\right)\check{Q}_{c}R_{s}^{2}.
\end{eqnarray}
Then the canonical variables can be expressed as
\begin{eqnarray}
\check{b}\left(r\right) & = & \gamma\sqrt{-\frac{\check{\lambda}\left(r\right)}{\check{\mu}\left(r\right)}},\\
\check{p}_{b}\left(r\right) & = & \frac{L_{0}}{r^{\text{sgn}\left(\beta_{b}\right)\check{Q}_{b}}}\sqrt{-\check{\lambda}\left(r\right)\check{\mu}\left(r\right)},\\
\check{c}\left(r\right) & = & -\frac{\gamma L_{0}R_{s}}{\sqrt{\check{\xi}\left(r\right)}},\\
\check{p}_{c}\left(r\right) & = & \frac{1}{2}\sqrt{\check{\xi}\left(r\right)}.
\end{eqnarray}
Replacing these expressions for the components of the extended metric (\ref{eq:metric-check})
yields the metric components
\begin{align}
\check{g}_{00}= & \frac{\check{p}_{b}^{2}}{L_{0}^{2}\check{p}_{c}}=-\frac{2}{r^{2\text{sgn}\left(\beta_{b}\right)\check{Q}_{b}}}\frac{\check{\lambda}\left(r\right)\check{\mu}\left(r\right)}{\sqrt{\check{\xi}\left(r\right)}},\label{eq:g00-check}\\
\check{g}_{11}= & -\frac{\gamma^{2}\check{p}_{c}}{\check{b}^{2}r^{2}}=\frac{1}{2r^{2}}\sqrt{\check{\xi}\left(r\right)}\frac{\check{\mu}\left(r\right)}{\check{\lambda}\left(r\right)},\label{eq:g11-check}\\
\check{g}_{22}= & \frac{\check{g}_{33}}{\sin^{2}\left(\theta\right)}=\frac{1}{2}\sqrt{\check{\xi}\left(r\right)}.\label{eq:g22-check}
\end{align}
Note that in order to get the correct signature for the metric, we
need $\check{\lambda}$ and $\check{\mu}$ to have the same sign which
means that, similar to some of the approaches in LQG \cite{Ashtekar:2020ckv},
$\check{b}$ and $\check{p}_{b}$ are imaginary in the exterior.

\subsection{Issues with the non-improved extended metric}

The full spacetime metric described by (\ref{eq:g00-check})-(\ref{eq:g22-check})
should have correct classical as well as asymptotic limits. It can be seen that although the classical limits are fine:
\begin{align}
\lim_{\beta_{b},\,\beta_{c}\to0}\bar{g}_{00} &= -\left(1-\frac{R_{s}}{r}\right),\\
\lim_{\beta_{b},\,\beta_{c}\to0}\bar{g}_{11} &= \left(1-\frac{R_{s}}{r}\right)^{-1},\\
\lim_{\beta_{b},\,\beta_{c}\to0}\bar{g}_{22} &= r^{2}.
\end{align}
The asymptotic limits are incorrect,
\begin{align}
\lim_{r\to\infty}\bar{g}_{00} &= \left\{\begin{array}{clc}
0, & \,\text{sgn}\left(\beta_{b}\right)=&1\\
-\infty, & \,\text{sgn}\left(\beta_{b}\right)=&-1
\end{array}\right.,\\
\lim_{r\to\infty}\bar{g}_{11} &= \left\{\begin{array}{clc}
\check{Q}_{b}, & \,\text{sgn}\left(\beta_{b}\right)=&1\\
1, & \,\text{sgn}\left(\beta_{b}\right)=&-1
\end{array}\right. .
\end{align}
Furthermore, the asymptotic expansion of the Kretschmann scalar
also falls off incorrectly as $K\propto r^{-4}$. This situation is similar to \cite{Ashtekar:2020ckv} where in addition to the issue with the fall-off of the Kretchmann scalar in the asymptotic region, this region is not maximally symmetric, and furthermore, e.g., surfaces that should be null in the asymptotic region, do not have this caudal structure \cite{Bouhmadi-Lopez:2019hpp}.

In LQG, certain prescriptions are applied to resolve similar issues
by making the quantum parameters of the models momentum-dependent
(or scale factor-dependent in Bianchi models)~\cite{Ashtekar:2006wn,Chiou:2012pg,Chiou:2008nm}.
These prescriptions are collectively known as improved schemes. Usually the term ``improved'' in this context refers to a scheme that leads to the correct classical and asymptotic limits and also removes large quantum gravity effects from low curvature regions. Inspired by these
methods, we will apply a similar prescription to the model at
hand. 

\section{Improved effective metric in the $\boldsymbol{\bar{\beta}}$ scheme \label{sec:Iimproved-metric}}

\subsection{Improved metric and its extension to the full spacetime\label{Sec:transit-to-improved-metric}}

As mentioned at the end of the previous section, we are now going
to make the quantum parameters $\beta_{b}$ and $\beta_{c}$ momentum-dependent. This is done by modifying the dimensionless quantum parameters
to
\begin{align}
\beta_{b}& \to  \bar{\beta}_{b}=\frac{\beta_{b}L_{0}^{4}}{p_{b}^{2}},\\
\beta_{c} &\to  \bar{\beta}_{c}=\frac{\beta_{c}L_{0}^{4}}{p_{c}^{2}},
\end{align}
where the powers of $L_{0}$ are included to render the $\bar{\beta}$'s
dimensionless. We call this prescription the $\bar{\beta}$ improved
scheme after a similar method used in LQG~\cite{Ashtekar:2006wn,Chiou:2012pg,Chiou:2008nm}. As a result of the above,
the effective GUP-induced algebra now becomes
\begin{align}
\left\{ b,p_{b}\right\} &=  G\gamma\left(1+\bar{\beta}_{b}b^{2}\right)=G\gamma\left(1+\frac{\beta_{b}L_{0}^{4}}{p_{b}^{2}}b^{2}\right),\label{eq:improved-algebra-b}\\
\left\{ c,p_{c}\right\} &=  2G\gamma\left(1+\bar{\beta}_{c}c^{2}\right)=2G\gamma\left(1+\frac{\beta_{c}L_{0}^{4}}{p_{c}^{2}}c^{2}\right).\label{eq:improved-algebra-c}
\end{align}
Note that this is done at the level of the Poisson algebra,
before solving the equations of motion in the interior. Hence, we
need to rework the entire interior dynamics with the same Hamiltonian
as (\ref{eq:H-tilde}), but using the improved algebra (\ref{eq:improved-algebra-b})-(\ref{eq:improved-algebra-c}).
As we will see, this modification remarkably resolves the asymptotic issues that
arose in the extension of the metric to the full spacetime in the
previous section, and furthermore leads to the singularity resolution in this black hole as we will see in the following sections. By singularity resolution we mean four things: finiteness of Riemann invariants everywhere in the spacetime, geodesic completeness, finiteness of both the expansion scalar and the Raychaudhuri equation, and nonvanishing of the radius of 2-spheres. We will present these results gradually in relevant sections in what follows.

Solving the equations of motion (see appendix \ref{app:Interior-EoM}) and fixing the integration constants
by matching the classical limits, $\beta_{b}\to0$ and $\beta_{c}\to0$,
of the canonical variables with the actual classical solutions (\ref{eq:class-sol-b-int-t})-(\ref{eq:class-sol-pc-int-t})
of the interior, results in 
\begin{align}
b &= \gamma\sqrt{\frac{-\left(\tilde{t}^{2}-\text{sgn}\left(\beta_{b}\right)Q_{b}\right)+R_{s}\sqrt{\tilde{t}^{2}-\text{sgn}\left(\beta_{b}\right)Q_{b}}}{\tilde{t}^{2}-\text{sgn}\left(\beta_{b}\right)Q_{b}}},\label{eq:b-int-extended}\\
p_{b} &= L_{0}\sqrt{\tilde{t}^{2}-\text{sgn}\left(\beta_{b}\right)Q_{b}}\sqrt{\frac{-\left(\tilde{t}^{2}-\text{sgn}\left(\beta_{b}\right)Q_{b}\right)+R_{s}\sqrt{\tilde{t}^{2}-\text{sgn}\left(\beta_{b}\right)Q_{b}}}{\tilde{t}^{2}-\text{sgn}\left(\beta_{b}\right)Q_{b}}},\\
c &= -\frac{\gamma L_{0}R_{s}}{2\left(\tilde{t}^{8}-\frac{1}{4}\text{sgn}\left(\beta_{c}\right)Q_{c}R_{s}^{2}\right)^{\frac{1}{4}}},\\
p_{c} &= \left(\tilde{t}^{8}-\frac{1}{4}\text{sgn}\left(\beta_{c}\right)Q_{c}R_{s}^{2}\right)^{\frac{1}{4}}.\label{eq:pc-int-extended}
\end{align}
In the above, we have defined two positive dimensionful quantum parameters
\begin{equation}
Q_{b} = \left|\beta_{b}\right|\gamma^{2} L_{0}^{2} \quad\textrm{and}\quad Q_{c} =  \left|\beta_{c}\right|\gamma^{2}L_{0}^{6},\label{eq:Qb-Qc-bar-imp-ext}
\end{equation}
where $Q_{b}$ has dimensions of $\text{[L]}^{2}$ and $Q_{c}$ has
dimensions of $\text{[L]}^{6}$. 

\begin{figure}[htb]
\begin{centering}
\includegraphics[scale=0.8]{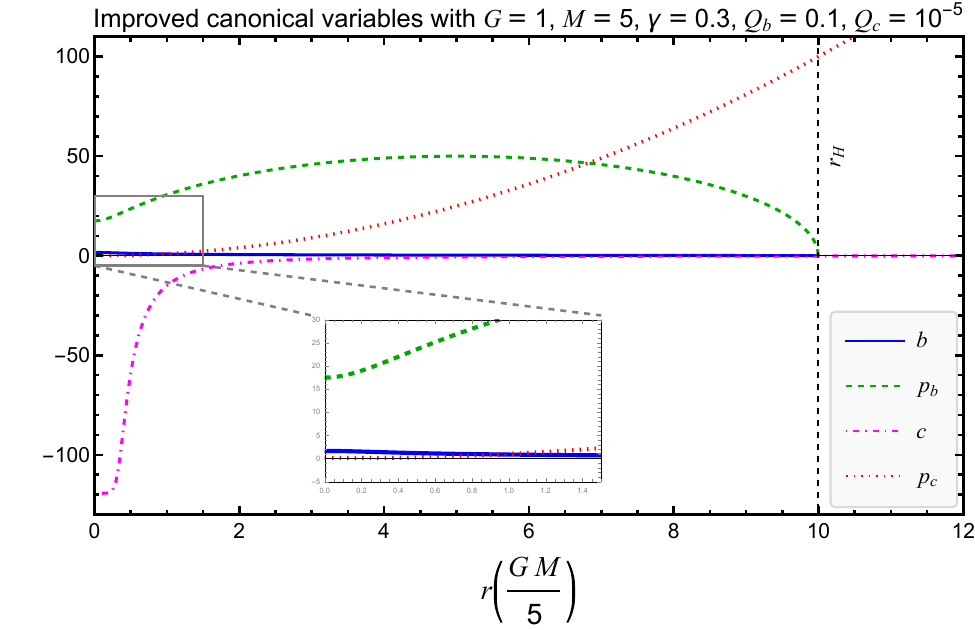}
\par\end{centering}
\caption{{\small{}Plot of improved canonical variables with the given values
of parameters on the top of the plot. Notice that both $b$ and $p_{b}$
become imaginary in the exterior. The horizon radius is $r_{H}$.
\label{fig:Canonical-Vars-Imp}}}
\end{figure}

Once again, we are going to postulate that, just as in the standard classical
Schwarzshild case, the full spacetime metric of this GUP-inspired
black hole is derived by analytical extension of the interior by switching
$\tilde{t}\to r,\,\tilde{r}\to t,\,\tilde{\theta}\to\theta$ and $\tilde{\phi}\to\phi$,
where $t$ and $r$ are the usual Schwarzshild coordinates. To obtain the
full spacetime metric in this way we make the aforementioned switch
of coordinate labels in (\ref{eq:b-int-extended})-(\ref{eq:pc-int-extended}),
and obtain the extended version of the canonical variables
as 
\begin{align}
b &= \gamma\sqrt{\frac{R_{s}}{\sqrt{\nu}}-1},\label{eq:b-ext-imp}\\
p_{b} &= L_{0}\sqrt{\nu}\sqrt{\frac{R_{s}}{\sqrt{\nu}}-1},\label{eq:pb-ext-imp}\\
c &= -\frac{\gamma L_{0}R_{s}}{2\rho^{\frac{1}{4}}},\label{eq:c-ext-imp}\\
p_{c} &= \rho^{\frac{1}{4}}.\label{eq:pc-ext-imp}
\end{align}
Here we have defined 
\begin{equation}
\nu = r^{2}-\text{sgn}\left(\beta_{b}\right)Q_{b} \quad\textrm{and}\quad \rho = r^{8}-\frac{1}{4}\text{sgn}\left(\beta_{c}\right)Q_{c}R_{s}^{2}.\label{eq:nu-rho-def-bar}
\end{equation}
Replacing the solutions (\ref{eq:b-ext-imp})-(\ref{eq:pc-ext-imp})
into the full spacetime metric
\begin{equation}
ds^{2}=-\frac{1}{r^{2}}\frac{\gamma^{2}p_{c}\left(r\right)}{b^{2}\left(r\right)}dr^{2}+\frac{p_{b}^{2}\left(r\right)}{L_{0}^{2}p_{c}\left(r\right)}dt^{2}+p_{c}\left(r\right)\left(d\theta^{2}+\sin^{2}\left(\theta\right)d\phi^{2}\right),\label{eq:met-int-r}
\end{equation}
we obtain the improved full metric components 
\begin{align}
g_{00} &= \frac{p_{b}^{2}}{L_{0}^{2}p_{c}}=-\sqrt{\frac{\nu}{\sqrt{\rho}}}\left(\sqrt{\nu}-R_{s}\right),\label{eq:g00-imp-ext-1}\\
g_{11} &= -\frac{\gamma^{2}p_{c}}{b^{2}r^{2}}=\frac{\sqrt{\nu\sqrt{\rho}}}{r^{2}\left(\sqrt{\nu}-R_{s}\right)},\label{eq:g11-imp-ext-1}\\
g_{22} &= \frac{g_{33}}{\sin^{2}\left(\theta\right)}=\rho^{\frac{1}{4}}.\label{eq:g22-imp-ext-1}
\end{align}
Before continuing, let us briefly examine the signature and the conditions
on reality of the above metric. To have a metric which
remains real for all values of coordinates, particularly for all values
of $r\in(0,\infty)$, we need both $\nu$ and $\rho$ to always remain
non-negative. As a result, from (\ref{eq:nu-rho-def-bar}) and (\ref{eq:Qb-Qc-bar-imp-ext})
we infer
\begin{align}
\nu  > 0 &\Rightarrow \text{sgn}\left(\beta_{b}\right)=-1,\label{eq:cond-sign-beta-b}\\
\rho > 0 &\Rightarrow \text{sgn}\left(\beta_{c}\right)=-1.\label{eq:cond-sign-beta-c}
\end{align}
This is the same result that had been obtained in previous works regarding
the interior of this black hole \cite{Blanchette:2021vid,Bosso:2020ztk,Rastgoo:2022mks,Bosso:2023fnb}.
This choice of the signs turns (\ref{eq:Qb-Qc-bar-imp-ext}) into
\begin{align}
Q_{b} &= \beta_{b}\gamma^{2}L_{0}^{2}\label{eq:Qb-ext-def}\, ,\\
Q_{c} &= \beta_{c}\gamma^{2}L_{0}^{6}\label{eq:Qc-ext-def}\, ,
\end{align}
and accordingly we rewrite (\ref{eq:nu-rho-def-bar}) as 
\begin{align}
\nu  &= r^{2}+Q_{b},\label{eq:nu-def}\\
\rho &= r^{8}+\frac{1}{4}Q_{c}R_{s}^{2}.\label{eq:rho-def}
\end{align}
With the conditions (\ref{eq:cond-sign-beta-b})-(\ref{eq:cond-sign-beta-c}),
not only the reality of the metric for all $r\in(0,\infty)$ is guaranteed,
but also its signature would be the expected one for both the interior
and the exterior. With these considerations, a plot of the canonical
variables (\ref{eq:b-ext-imp})-(\ref{eq:pc-ext-imp}) 
is shown in Fig.~\ref{fig:Canonical-Vars-Imp}. From \eqref{eq:g22-imp-ext-1}
and \eqref{eq:rho-def} (and Fig.~\ref{fig:Canonical-Vars-Imp}),
one can see as $r\to0$, the square of the radius
of 2-spheres $g_{22}=p_{c}$ will not vanish. Although not a proof, but given the form of the Kretchmann scalar \eqref{K-Classic} with $p_c$ in the denominator, this already
signals we are moving in the right direction with regard to the resolution of the classical singularity. 

\begin{figure}[htb]
\begin{centering}
\includegraphics[scale=0.8]{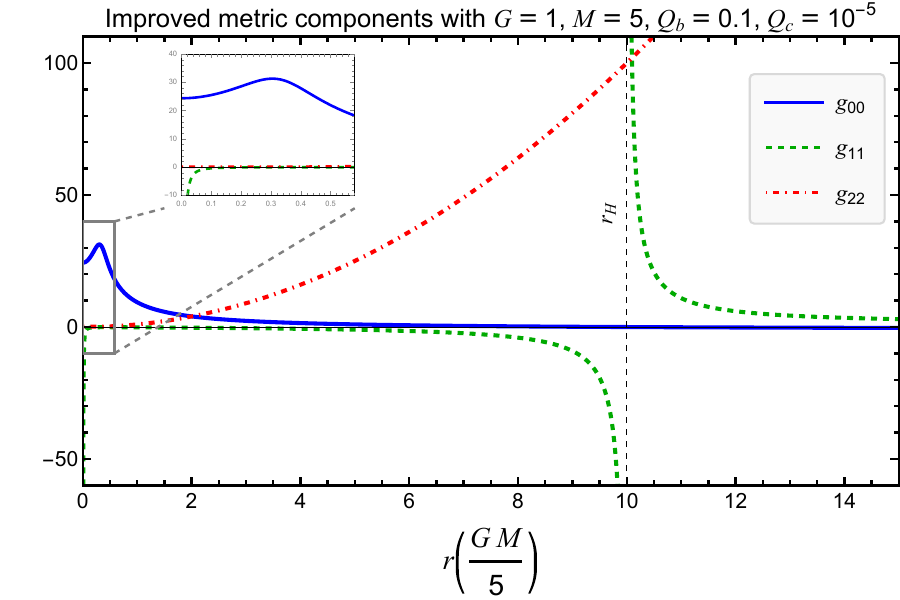}
\par\end{centering}
\caption{{\small{}Plot of improved metric components in Schwarzschild coordinates
with the given values of parameters on the top of the plot. Note the divergence of $g_{11}$ at $r=0$.}}
\label{fig:metric-Imp}
\end{figure}

Hence, the final expression for the improved full spacetime metric
components are (\ref{eq:g00-imp-ext-1})-(\ref{eq:g22-imp-ext-1})
together with (\ref{eq:nu-def})-(\ref{eq:rho-def}), and (\ref{eq:Qb-ext-def})-(\ref{eq:Qc-ext-def}),
which are explicitly written as
\begin{align}
g_{00} &= -\sqrt{\frac{\nu}{\sqrt{\rho}}}\left(\sqrt{\nu}-R_{s}\right)\nonumber \\
&= -\left(1+\frac{Q_{b}}{r^{2}}\right)\left(1+\frac{Q_{c}R_{s}^{2}}{4r^{8}}\right)^{-1/4} 
\left(1-\frac{R_{s}}{\sqrt{r^2+Q_b}}\right)
,\label{eq:g00-imp-ext}\\
g_{11} &= \frac{\sqrt{\nu\sqrt{\rho}}}{r^{2}\left(\sqrt{\nu}-R_{s}\right)}= 
\left(1+\frac{Q_{c}R_{s}^{2}}{4r^{8}}\right)^{1/4}
\left(1-\frac{R_{s}}{\sqrt{r^2+Q_b}}\right)^{-1}
,\label{eq:g11-imp-ext}\\
g_{22} &=\frac{g_{33}}{\sin^{2}\left(\theta\right)} = \rho^{\frac{1}{4}}=r^{2}\left(1+\frac{Q_{c}R_{s}^{2}}{4r^{8}}\right)^{1/4}.\label{eq:g22-imp-ext}
\end{align}
The coordinates have the natural domain $r \in [0,+\infty)$, $t \in (-\infty,+\infty)$, $\theta \in [0,\pi]$ and $\phi \in [0,2\pi]$. A plot of these metric components are presented in Fig.~\ref{fig:metric-Imp}.
The existence of the minimum radius of 2-spheres is related to the resolution of the singularity and we see that the quantum effects close to the singularity are associated to $\beta_{c}$ as expected. 

The above metric components depend on three real non-zero positive parameters: $M$, $Q_b$ and $Q_c$. As we will see later in detail in Sec. \ref{sec:horizon}, certain relations between these three parameters lead to three types of spacetimes: a black hole, a wormhole, or a remnant.

Before moving on to the properties of this metric we would like to briefly discuss the consequences of Eqs. \eqref{eq:cond-sign-beta-b} and \eqref{eq:cond-sign-beta-c}. The negativity of $\beta_{b},\,\beta_{c}$ does not allow the existence
of minimal uncertainty. This means that the theory is in the regime
of generalized uncertainty instead. As discussed in \cite{Jizba:2009qf},
if such a  model is considered in the realm of quantum mechanics where $X$ is the position operator with eigenvalues $x$, and $P$ is the momentum operator with eigenvalues $p$, and a version of generalized uncertainty is used where $p$ is on the right hand side  of the commutators multiplied by $\beta$, then this leads to a crystal-like universe where there is a lattice in $x$ with a lattice spacing of the order of Planck length. Furthermore, for energies that
are near the border of the Brillouin zone, in this case Planckian
energies, the uncertainty relation for position and momenta does not
pose any lower bound on uncertainties.

In our case, there are two main differences from the above treatment: 1) our algebra is not on points of space $x$ and their momenta, since in general relativity these points have no physical meaning due to diffeomorphism invariance. Instead, the fundamental algebra that is modified is the algebra of the metric (or equivalently triads) and its conjugate. 2) In our model, the variable on the right hand side of the Poisson bracket multiplied by $\bar{\beta}$'s is not the momentum but the configuration variable (components of the Ashtekar-Barbero connection). One then expects that the lattice behavior is now associated to the triads or the metric. This very much makes sense since if triads can only take certain discrete values, one is dealing with a discrete metric and consequently a quantum spacetime.  

\subsection{Classical and asymptotic limits }

It is quite easy to see that the classical limit $\beta_{b}\to0$ and $\beta_{c}\to0$
(or equivalently $Q_{b}\to0$ and $Q_{c}\to0$ according to (\ref{eq:Qb-ext-def})-(\ref{eq:Qc-ext-def}))
of the improved metric components (\ref{eq:g00-imp-ext})-(\ref{eq:g22-imp-ext})
match those of the classical Schwarzschild metric; explicitly 
\begin{align}
\lim_{\beta_{b},\,\beta_{c}\to0}g_{00} &= -\left(1-\frac{R_{s}}{r}\right), & \lim_{\beta_{b},\,\beta_{c}\to0}g_{11} &= \left(1-\frac{R_{s}}{r}\right)^{-1}, & \lim_{\beta_{b},\,\beta_{c}\to0}g_{22} &= r^{2}.
\end{align}
For the asymptotic limit, we have 
\begin{align}
\lim_{r\to\infty}g_{00} &= -1, & \lim_{r\to\infty}g_{11} & = 1, & \lim_{r\to\infty}g_{22} &= \infty.
\end{align}
This improved prescription fixes the issues with the
asymptotic limit and the asymptotic expansion of the metric. The asymptotic expansions also match the Schwarzschild spacetime to leading order: 
\begin{align}
g_{00}\big|_{r\to\infty} &= -\left(1-\frac{R_{s}}{r}\right)-\frac{Q_{b}}{r^{2}}+\mathcal{O}\left(\frac{1}{r}\right)^3\label{eq:g00-imp-ext-series-inf},\\
g_{11}\big|_{r\to\infty} &= \left(1+\frac{R_{s}}{r}\right)+\frac{R_{s}^{2}}{r^{2}}+\frac{R_{s}}{2r^{3}}\left(2R_{s}^{2}-Q_{b}\right)+\mathcal{O}\left(\frac{1}{r}\right)^4\label{eq:g11-imp-ext-series-inf},\\
g_{22}\big|_{r\to\infty} &=r^{2}+\frac{Q_{c}R_{s}^{2}}{16r^{6}}+\mathcal{O}\left(\frac{1}{r}\right)^{14}.\label{eq:g22-imp-ext-series-inf}
\end{align}
Notice that as expected from previous studies, the asymptotic and
large $r$ behavior is governed by $\beta_{b}$. On the other hand
the singularity resolution is governed by $\beta_{c}$ as we will
see in the next section. 


\subsection{Spacetime structure, horizon and  mass\label{sec:horizon}}

Since the model is static and spherically symmetric, the event horizon is a Killing horizon, and we can obtain the
position of the event horizon simply by solving either $g^{11}(r_H)=0$ (for event horizons) or $g_{00}(r_H)=0$ (for Killing horizons corresponding to the asymptotic timelike Killing vector field). This
yields the horizon radius as
\begin{equation}
r_{H}=\sqrt{R_{s}^{2}-Q_{b}}=R_{s}\sqrt{1-\frac{Q_{b}}{R_{s}^{2}}},\label{eq:r-Horiz}
\end{equation}
which is smaller than the Schwarzschild radius, albeit by a very small
amount. As should be clear, this is a pure quantum effect. Up to the
first order in $Q_{b}$ we have
\begin{equation}
r_{H}=R_{s}-\frac{1}{2}\frac{Q_{b}}{R_{s}}+\mathcal{O}\left(\frac{Q_{b}^{2}}{R_s^3}\right).
\end{equation}
Given the solution for the horizon, different spacetimes
are possible depending on the relative values of $M$ and $Q_{b}$.
To see this better, let us first make a transformation 
\begin{equation}
\bar{r}^{8}=r^{8}+\frac{1}{4}Q_{c}R_{s}^{2}
\end{equation}
such that in the new coordinate system $g_{22}=\bar{r}^{2}$. Notice
that now the minimum radius of 2-spheres is
\begin{equation}
\bar{r}_{\text{min}}=\bar{r}(r=0)=\left(\frac{1}{4}Q_{c}R_{s}^{2}\right)^{\frac{1}{8}}.
\end{equation}
Furthermore, in this new coordinate system, the position of the horizon
becomes
\begin{align}
\bar{r}_{H}^{8}= & r_{H}^{8}+\frac{1}{4}Q_{c}R_{s}^{2}\nonumber\\
= & \left(R_{s}^{2}-Q_{b}\right)^{4}+\frac{1}{4}Q_{c}R_{s}^{2},
\end{align}
where we have used (\ref{eq:r-Horiz}). Now we will have three cases.
For $\bar{r}_{H}>\bar{r}_{\text{min}}$, or equivalently 
\begin{equation}
\left(R_{s}^{2}-Q_{b}\right)^{4}>0\Rightarrow M>\frac{\sqrt{Q_{b}}}{2G},
\end{equation}
the spacetime exhibits an event horizon (two solutions for $g^{11}=0$,
one with a positive $r$ value and one with a negative $r$ value, or the equivalent in the $\bar{r}$ coordinate system)
and is thus a black hole. This is confirmed by an analysis of the trapped region of the spacetime (see Eq. \eqref{theta_pm_Horiz} and the paragraph after that). The extremal case of $\bar{r}_{H}=\bar{r}_{\text{min}}$,
or equivalently
\begin{equation}
\left(R_{s}^{2}-Q_{b}\right)^{4}=0\Rightarrow M=\frac{\sqrt{Q_{b}}}{2G},
\end{equation}
is a natural limit on the minimum black hole mass. A stability analysis
is needed to determine if the black hole remnant is stable. Finally
the case of $\bar{r}_{H}<\bar{r}_{\text{min}}$, or equivalently 
\begin{equation}
\left(R_{s}^{2}-Q_{b}\right)^{4}<0\Rightarrow M<\frac{\sqrt{Q_{b}}}{2G},
\end{equation}
where the horizon radius is smaller than the minimum radius of 2-spheres. If it is possible, it would describe a spacetime
with no event horizon, where due to the presence of a minimum radius
of 2-spheres, it would be a one-way wormhole with an extremal
null throat at $\bar{r}=\bar{r}_{\text{min}}$~\cite{Simpson:2018tsi}.

To gain insight into the meaning of the $M$ parameter, we calculate
the usual geometrical definitions of mass. We first calculate the
Komar mass~\cite{PhysRev.129.1873} starting from the the surface
integral expression~\cite{Carroll:book}.

\begin{equation}
M_{\text{K}}(r)=\frac{1}{4\pi G}\int_{\partial\Sigma}d^{2}x\sqrt{\gamma^{(2)}}\ n_{\mu}\sigma_{\nu}\nabla^{\mu}K^{\nu}\,,
\end{equation}
where $n^{\mu}$ is a unit normal timelike vector to the spacelike
hypersurface $\Sigma$ with constant $t$, and $\sigma^{\mu}$ is
a unit spacelike normal vector to the 2-spheres which is the boundary
of $\partial\Sigma$ at infinity, $\gamma_{ij}^{(2)}$ is the metric
of that 2-sphere, and $K^{\nu}$ is the timelike Killing vector of
the static spacetime. A unit normal vector $v^{\mu}$ to a hypersurface
described by a function $f(x)=C$ with $C$ a constant can be written
as
\begin{equation}
v^{\mu}=\pm\sqrt{\frac{\pm1}{g^{\alpha\beta}\partial_{\alpha}f\partial_{\beta}f}}g^{\mu\nu}\partial_{\nu}f\, ,
\end{equation}
where the positive signs correspond to a spatial unit vector and negative
signs correspond to a timelike unit vector. With a diagonal metric, for a spacelike unit vector
$\sigma^{\mu}$ normal to a surface described by $r=C_{1}$,
we obtain 
\begin{equation}
\sigma^{\mu}=\frac{g^{\mu r}}{\sqrt{g^{rr}}}\, ,
\end{equation}
while for a timelike unit vector normal to a hypersurface $t=C_{2}$,
we get
\begin{equation}
n^{\mu}=-\frac{g^{\mu t}}{\sqrt{-g^{tt}}}\, .
\end{equation}
Replacing these together with $K^{\mu}=(1,0,0,0)$ and $\sqrt{\gamma^{(2)}}=g_{\theta\theta}\sin(\theta)$
inside $M_{\text{K}}$ yields
\begin{align}
M_{\text{K}}= & \frac{r^{2}}{2G}\left(\frac{1}{\sqrt{\nu}}-\frac{\omega}{\nu}+2r^{6}\frac{\omega}{\rho}\right)\nonumber \\
= & \frac{r^{2}}{2G}\left(\frac{1}{\sqrt{r^{2}+Q_{b}}}-\frac{R_{s}-\sqrt{r^{2}+Q_{b}}}{r^{2}+Q_{b}}+2r^{6}\frac{R_{s}-\sqrt{r^{2}+Q_{b}}}{r^{8}+\frac{1}{4}Q_{c}R_{s}^{2}}\right)\, ,
\end{align}
where we have defined 
\begin{equation}
\omega=R_{s}-\sqrt{\nu}=R_{s}-\sqrt{r^{2}+Q_{b}}\, .\label{eq:omega-def}
\end{equation}
The asymptotic expansion of $M_{\text{K}}$ as $r\to\infty$ is 
\begin{equation}
M_{\mathrm{K}}(r)=M-\frac{Q_{b}}{Gr}+\mathcal{O}\left(\frac{1}{r^{2}}\right),
\end{equation}
from which it is seen that the asymptotic limit of the Kumar
mass is actually the parameter $M$ in the metric,
\begin{equation}
\lim_{r\to\infty}M_{\mathrm{K}}(r)=M.
\end{equation}

Next, we compute the ADM mass~\cite{PhysRev.116.1322} using the
expression~\cite{Gourgoulhon:2007ue}

\begin{equation}
M_{\text{ADM}}=\lim_{r\to\infty}\bar{M}_{\text{ADM}}=\lim_{r\to\infty}\frac{1}{16\pi G}\int_{\partial\Sigma}d^{2}x\sqrt{\gamma^{(2)}}\sigma^{i}\left(\bar{\eta}^{jk}\mathcal{D}_{k}q_{ij}-\mathcal{D}_{i}\left(\bar{\eta}^{jk}q_{jk}\right)\right)\,,
\end{equation}
where $\mathcal{D}$ is the covariant derivative associated with the
background metric $\bar{\eta}^{jk}$ which is the flat metric in spherical
coordinates in our case, $\sigma^{i}=\bar{\eta}^{ir}/\sqrt{\bar{\eta}^{rr}}=(1,0,0)$
is the unit spacelike normal outward-pointing vector to the 2-sphere
$\partial\Sigma$ at infinity, which is the boundary of the spatial
hypersurface $\Sigma$. Also $q_{ij}$ is the induced
metric on the 3D spatial hypersurface $\Sigma$, and $\gamma_{ij}$
is the metric of the 2-sphere $\partial\Sigma$. Using these, we obtain
\begin{align}
\bar{M}_{\text{ADM}}=~ & \frac{\omega\left(\rho-2r^{8}\right)-\sqrt{\nu}\rho}{2G\sqrt{\rho}r^{3}\omega}\nonumber \\
=~ & \frac{1}{2Gr^{3}}\left[\frac{\frac{1}{4}Q_{c}R_{s}^{2}-r^{8}}{\sqrt{r^{8}+\frac{1}{4}Q_{c}R_{s}^{2}}}-\frac{\sqrt{r^{2}+Q_{b}}\sqrt{r^{8}+\frac{1}{4}Q_{c}R_{s}^{2}}}{R_{s}-\sqrt{r^{2}+Q_{b}}}\right].
\end{align}
The asymptotic expansion for $r\to\infty$ then reads 
\begin{equation}
\bar{M}_{\mathrm{ADM}}=M+\frac{2GM^{2}}{r}+\frac{M\left(8G^{2}M^{2}-Q_{b}\right)}{2r^{2}}+\mathcal{O}\left(\frac{1}{r^{3}}\right),
\end{equation}
from which we can see
\begin{equation}
M_{\text{ADM}}=\lim_{r\to\infty}\bar{M}_{\text{ADM}}=M.
\end{equation}
Finally we turn our attention to the Hawking quasilocal mass, which for spherical symmetry reduces to the
Misner-Sharp-Hernandez (MSH) mass~\cite{PhysRev.136.B571,Faraoni:2020mdf} given by
\begin{align}
M_{\text{MSH}}(r)&= \frac{r}{2G}(1-\nabla^{a}r\nabla_{a}r)=  \frac{r}{2G}\left(1-\frac{1}{g_{11}}\right)\nonumber \\
&= \frac{r}{2G}\left(1-\left(1-\frac{R_{s}}{\sqrt{r^{2}+Q_{b}}}\right)\left(1+\frac{Q_{c}R_{s}^{2}}{4r^{8}}\right)^{-1/4}\right).\label{eq:MMSH}
\end{align}
It is also evident that  $M_\mathrm{MSH}(r)\to M$ as $r\to\infty$.

We thus find
that the parameter $M$ in our GUP-inspired black hole is the same as the above three masses
\begin{equation}
\lim_{r\to\infty} \bar{M}_{\text{ADM}}(r)
=\lim_{r\to\infty} M_{\text{MSH}}(r)
=\lim_{r\to\infty} M_{\text{K}}(r) =M.
\end{equation}

\subsection{Kretschmann scalar}

\begin{figure}
\begin{centering}
\includegraphics[scale=0.8]{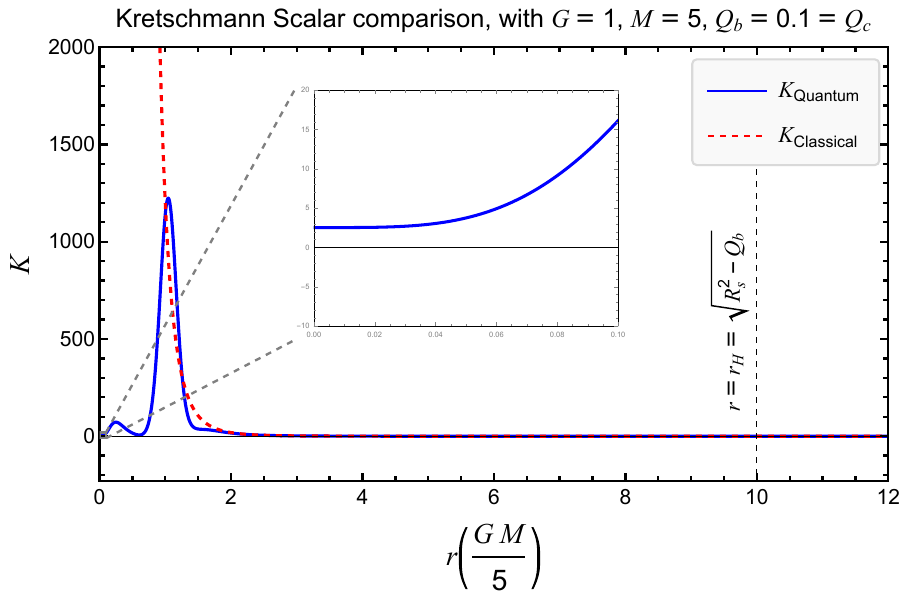}
\par\end{centering}
\caption{{\small{}Comparing the classical and effective Kretschmann scalars.
In this figure, we have zoomed in on the region where the classical
and quantum Kretschmann scalars start to deviate. Notice that a different value of $Q_c$ is used from that in previous plots.\label{fig:Kretsch-1}}}
\end{figure}

In addition to the metric, we need to also make sure that the
asymptotic limit and asymptotic expansion of the Kretschmann scalar
$K$ matches those of the classical Schwarzschild spacetime. The expression
for $K$ in terms of the canonical variables is found to be
\begin{align}
K= & \frac{4}{\sqrt{\rho}}\nonumber \\
 & +\frac{4r^{4}}{\nu^{5}\rho^{\frac{9}{2}}}\left\{ \rho^{4}\left(\nu^{3/2}-\nu\omega\right)^{2}+2\nu\rho^{4}\omega\left(\sqrt{\nu}-\omega\right)r^{2}+\rho^{4}\omega^{2}r^{4}\right.\nonumber\\
 & +16\nu^{3}\rho^{3}\omega\left(\sqrt{\nu}-\omega\right)r^{6}+2\nu^{2}\rho^{3}\left(3\sqrt{\nu}\omega-2\nu+7\omega^{2}\right)r^{8}\nonumber \\
 & +2\nu\rho^{3}\omega\left(\omega-2\sqrt{\nu}\right)r^{10}+2\nu^{4}\rho^{2}\omega\left(\sqrt{\nu}+96\omega\right)r^{12}\nonumber \\
 & +4\nu^{3}\rho^{2}\omega\left(5\omega-17\sqrt{\nu}\right)r^{14}+\nu^{2}\rho^{2}\left(-4\sqrt{\nu}\omega+5\nu-18\omega^{2}\right)r^{16}\nonumber \\
 & \left.-416\nu^{4}\rho\omega^{2}r^{20}+2\nu^{3}\rho\omega\left(29\sqrt{\nu}-3\omega\right)r^{22}+231\nu^{4}\omega^{2}r^{28}\right\} ,\label{eq:K-expression}
\end{align}
where we have used \eqref{eq:nu-def}, \eqref{eq:rho-def} and \eqref{eq:omega-def}. 
The Kretschmann expression (\ref{eq:K-expression}) has all the desired
properties. First, the asymptotic limit of the Kretschmann scalar
vanishes,
\begin{equation}
\lim_{r\to\infty}K=0.
\end{equation}
Second, the asymptotic expansion is 
\begin{equation}
K\big|_{r\to\infty}=\frac{12R_{s}^{2}}{r^{6}}+\mathcal{O}\left(\frac{1}{r}\right)^7,
\end{equation}
in which the leading term is precisely the Schwarzschild Kretschmann
scalar. Third, $K$ is always finite over the entire spacetime 
$r\in[0,\infty]$. 
In fact noting 
\begin{align}
\nu\left(r=0\right)=\lim_{r\to0^{+}}\nu &= Q_{b},\label{eq:nu-at-zero}\\
\rho\left(r=0\right)=\lim_{r\to0^{+}}\rho &= \frac{1}{4}Q_{c}R_{s}^{2},\label{eq:rho-at-zero}\\
\omega\left(r=0\right)=\lim_{r\to0^{+}}\omega &= R_{s}-\sqrt{\nu\left(r=0\right)}=R_{s}-\sqrt{Q_{b}},\label{eq:omega-at-zero}
\end{align}
it is rather easy to see from (\ref{eq:K-expression}) that at $r=0$,
which previously exhibited a singularity,
we encounter a finite value for the Kretschmann scalar,
\begin{equation}
\lim_{r\to0^{+}}K=K\left(r=0\right)=\frac{4}{\sqrt{\rho}}\bigg|_{r=0}=\frac{8}{R_{s}\sqrt{Q_{c}}}.
\end{equation}
The finiteness of $K(r=0)$ was already observed in previous works
that only considered the interior \cite{Blanchette:2021vid,Bosso:2020ztk,Rastgoo:2022mks,Bosso:2023fnb}.
This regularity of $K$ is one of the evidences supporting the claim that the singularity of the black
hole is resolved in this effective model. The fourth and final property
of the Kretschmann expression (\ref{eq:K-expression}), given its
denominator and noting the definitions of $\nu$ and $\rho$, ((\ref{eq:nu-def})
and (\ref{eq:rho-def}), respectively), is that $K$ is regular also at the horizon
$r=r_{H}=\sqrt{R_{s}^{2}-Q_{b}}$, which is another important and
desired property. This can be seen explicitly from the values of $\nu$
and $\rho$ at the horizon,
\begin{align}
\nu\left(r=r_{H}\right) &= R_{s}^{2},\\
\rho\nu\left(r=r_{H}\right)= & \left(R_{s}^{2}-Q_{b}\right)^{4}+\frac{1}{4}Q_{c}R_{s}^{2}.
\end{align}
Some of the above properties can be directly seen from the plot of
the Kretschmann scalar in Fig. \ref{fig:Kretsch-1}. 

Note that these finite and regular values at the horizon and at $r=0$
and also the correct asymptotic behavior, are a direct consequence of
choosing the signs of the $\beta$'s in (\ref{eq:cond-sign-beta-b})-(\ref{eq:cond-sign-beta-c}),
which themselves are the result of the reality condition for the metric.

For completeness, let us also calculate the expansion of the Ricci scalar and Ricci tensor squared in the asymptotic as well as quantum regimes.
The asymptotic expressions at spatial infinity read 
\begin{equation}
g^{\mu\nu} R_{\mu\nu}   = -\frac{2Q_b}{r^4} + \mathcal{O}\left(\frac{1}{r}\right)^5 \quad \textrm{and} \quad    
R_{\mu\nu} R^{\mu\nu} = \frac{12Q_b}{r^8} + \mathcal{O}\left(\frac{1}{r}\right)^9\, .    
\end{equation}
In the quantum regime, as $r\to 0$, 
\begin{equation}
g^{\mu\nu}R_{\mu\nu} = \frac{2}{(Q_c R_s^2/4)^{1/4}} 
\quad \textrm{and}\quad
R_{\mu\nu} R^{\mu\nu} = \frac{2}{(Q_c R_s^2/4)^{1/2}} \, .    
\end{equation}
It is seen that the main deviations from general relativity are dominated by the quantum parameter $Q_b$ and the scalars at $r=0$ are finite, depending on the $Q_c$ quantum parameter.

\subsection{Effective stress-energy tensor \label{sec:Effective-stress-energy-tensor}}

The properties of the effective quantum geometry can be visualized
by an effective stress-energy tensor 
\begin{equation}
T_{\mu\nu}=\frac{1}{8\pi G}G_{\mu\nu}\,,
\end{equation}
where $G_{\mu\nu}$ is the Einstein tensor. Let us start by considering
a stress-energy tensor that is in the form of an anisotropic perfect
fluid 
\begin{equation}
T^{\mu\nu}=\left(\epsilon+p_{\theta}\right)U^{\mu}U^{\nu}+\left(p_{r}-p_{\theta}\right)W^{\mu}W^{\nu}+p_{\theta}g^{\mu\nu},
\end{equation}
characterized by an effective energy density $\epsilon$, and radial
and tangential pressure densities $p_{r}$ and $p_{\theta}$ $(=p_{\phi})$,
respectively. Here $U^{\mu}$ is a unit timelike vector field, $U^{\mu}U_{\mu}=-1$,
which for a comoving fluid becomes 
\begin{equation}
U^{\mu}=\left(\sqrt{-g^{00}},0,0,0\right),
\end{equation}
and $W^{\mu}$ is a unit spacelike vector field, $W^{\mu}W_{\mu}=1$,
and satisfying $U^{\mu}W_{\mu}=0$. With these conditions, it becomes 
\begin{equation}
W^{\mu}=\left(0,\sqrt{g^{11}},0,0\right).
\end{equation}
Using Einstein's equation we can then write for the exterior
\begin{align}
\epsilon= & -\frac{1}{8\pi G}g^{00}G_{00},\\
p_{r}= & \frac{1}{8\pi G}g^{11}G_{11},\\
p_{\theta}= & \frac{1}{8\pi G}g^{22}G_{22}.
\end{align}
For the interior, we need to switch $g^{00}\leftrightarrow g^{11}$ and $G_{00}\leftrightarrow G_{11}$, and hence $\epsilon\leftrightarrow -p_r $.
Thus, in the quantum region, $r=0$, we get 
\begin{align}
\epsilon & =\frac{1}{4\sqrt{2}\pi G}\frac{1}{\left(Q_{c}R_{s}^{2}\right)^{1/4}},\label{eq:E-density-0}\\
p_{r} & =-\frac{1}{4\sqrt{2}\pi G}\frac{1}{\left(Q_{c}R_{s}^{2}\right)^{1/4}},\label{eq:pr-density-0}\\
p_{\theta} & =0,\label{eq:ptheta-density-0}
\end{align}
Furthermore, the asymptotic behavior of these quantities at $r\to\infty$ 
is 
\begin{align}
\epsilon= & \frac{1}{8\pi G}\frac{R_{s}Q_{b}}{r^{5}}+\mathcal{O}\left(\frac{1}{r}\right)^{7}\,,\label{eq:E-density-inf}\\
p_{r}= & -\frac{1}{4\pi G}\frac{Q_{b}}{r^{4}}+\mathcal{O}\left(\frac{1}{r}\right)^{5}\,,\label{eq:pr-density-inf}\\
p_{\theta}= & \frac{1}{4\pi G}\frac{Q_{b}}{r^{4}}+\mathcal{O}\left(\frac{1}{r}\right)^{5}\,.\label{eq:ptheta-density-inf}
\end{align}
The fall-off is as rapid as $r^{-4}$ in which the Schwarzschild spacetime
is recovered. The asymptotic behavior is dominated by $Q_{b}$ and
corrections due to $Q_{c}$ are subdominant.

We can now consider the energy conditions. The weak energy condition is
satisfied if $\epsilon\geq0$ and $\epsilon+p_{i}\geq0$ with $i=r,\,\theta,\,\phi$,
noting that $p_{\theta}=p_{\phi}$. The strong energy condition is
satisfied if $\epsilon+p_{i}\geq0$ and $\epsilon+\sum_{i}p_{i}\geq0$.
Finally, the dominant energy condition is satisfied if $\epsilon\geq|p_i|$. From \eqref{eq:E-density-0}-\eqref{eq:ptheta-density-0}, we see that at the origin, $r=0$, the weak, strong and dominant energy conditions are satisfied. However, at $r\to\infty$, the leading terms of  \eqref{eq:E-density-inf}-\eqref{eq:ptheta-density-inf}, violate all energy conditions, although they all vanish at infinity and thus are nonviolating.

We would like to emphasize that these are not (non)violation
of an actual matter field energy condition, since this is a vacuum
model. One only obtains an stress-energy tensor if one uses the Einstein's
equations. However, these are not the equations of motion in our model,
since this is a modified theory due to quantum gravity effects. So
one should not read too much into the above energy conditions. We
have only presented the above results for completeness.


\section{Geodesics \label{sec:Geodesics}}


Since the metric is static and spherically symmetric, there
are three spatial Killing vector fields and one asymptotically timelike
one, associated to the rotational symmetry of geodesics and their energy
in this spacetime, respectively. Hence, we can fix the direction of
the angular momentum ($\theta=\pi/2$) and write
its conserved magnitude as 
\begin{equation}
L=g_{22}\frac{d\phi}{d\lambda}=\rho^{\frac{1}{4}}\frac{d\phi}{d\lambda},
\end{equation}
where $\lambda$ is the affine parameter of the geodesics. In the
same way, the conserved energy of the geodesic is 
\begin{equation}
E=-g_{00}\frac{dt}{d\lambda}=\sqrt{\frac{\nu}{\sqrt{\rho}}}\left(\sqrt{\nu}-R_{s}\right)\frac{dt}{d\lambda}.
\end{equation}
The conservation of the velocity vector (momentum for null vector)
of the geodesic yields
\begin{equation}
\varepsilon=-g_{\mu\nu}\frac{dx^{\mu}}{d\lambda}\frac{dx^{\nu}}{d\lambda}=\begin{cases}
0, & \text{null geodesics}\\
1, & \text{timelike geodesics}
\end{cases}.
\end{equation}
Combining these conserved quantities, we obtain in general,
\begin{equation}
-\frac{1}{2}g_{00}g_{11}\left(\frac{dr}{d\lambda}\right)^{2}-\frac{1}{2}g_{00}\left[\frac{L^{2}}{g_{22}}+\varepsilon\right]=\frac{1}{2}E^{2},\label{eq:geod-equiv-1}
\end{equation}
which can be written more explicitly in our model as
\begin{equation}
\frac{1}{2}\frac{\nu}{r^{2}}\left(\frac{dr}{d\lambda}\right)^{2}+V_{\text{eff}}=\bar{E},
\end{equation}
such that the energy is not $r$ dependent.
We have defined the effective potential $V_{\text{eff}}$ and
the energy $\bar{E}$ as
\begin{align}
V_{\text{eff}} &= -\frac{1}{2}g_{00}\left[\frac{L^{2}}{g_{22}}+\varepsilon\right]=\frac{1}{2}\sqrt{\frac{\nu}{\sqrt{\rho}}}\left(\sqrt{\nu}-R_{s}\right)\left[\frac{L^{2}}{\rho^{\frac{1}{4}}}+\varepsilon\right],\label{eq:Veff-gen}\\
\bar{E} &= \frac{1}{2}E^{2}.
\end{align}
Given that $\nu/r^{2}$ never vanishes in our model, the
circular orbits are located at $r$ for which $dV_{\text{eff}}/dr=0$.
Using (\ref{eq:Veff-gen}) in terms of a general $g_{00}$
and $g_{22}$, this condition is translated to
\begin{equation}
\varepsilon g_{22}^{2}\frac{dg_{00}}{dr}+L^{2}\left(g_{22}\frac{dg_{00}}{dr}-g_{00}\frac{dg_{22}}{dr}\right)=0,\label{eq:circular-orbit-condition}
\end{equation}
for a general stationary spherically symmetric metric in diagonal
form.

\subsection{Null geodesics and the photon sphere}

\begin{figure}
\begin{centering}
\includegraphics[scale=0.8]{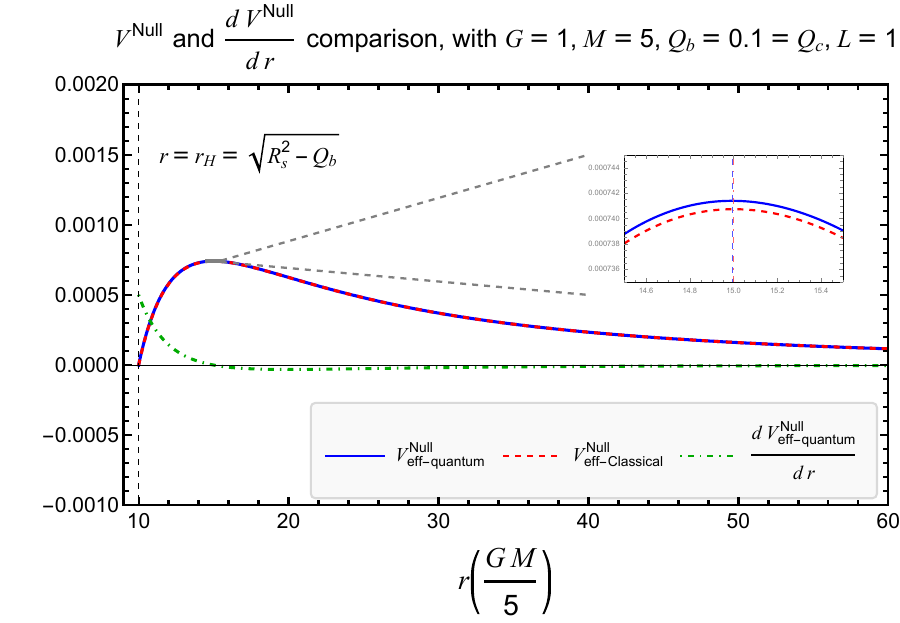}
\par\end{centering}
\caption{{\small{}The null effective potential for both the classical and quantum
cases, and the derivative of the effective potential in the exterior.
It is seen that there seems to be only one extremum in the exterior
which is a maximum. This marks the position of the photon sphere.
The zoomed-in part shows that the effective photon sphere has a smaller
radius (vertical dashed blue line) compared to the classical one (vertical
dot dashed red line at $3R_{2}/2$). \label{fig:Veff-Null-Exter}}}
\end{figure}

For null geodesics $\varepsilon=0$ and the condition for the
photon sphere(s) from \eqref{eq:Veff-gen} and (\ref{eq:circular-orbit-condition}), we have
\begin{equation}
V_{\text{eff}} = -\frac{1}{2}g_{00}\left[\frac{L^{2}}{g_{22}}\right]=\frac{1}{2}\sqrt{\frac{\nu}{\sqrt{\rho}}}\left(\sqrt{\nu}-R_{s}\right)\left[\frac{L^{2}}{\rho^{\frac{1}{4}}}\right],
\end{equation}
\begin{equation}
g_{22}\frac{dg_{00}}{dr}-g_{00}\frac{dg_{22}}{dr}=0.
\end{equation}
In our model, this last condition yields
\begin{equation}
\frac{1}{2}\left(1+\sqrt{\rho}\right)\frac{1}{\rho}\frac{d\rho}{dr}=\left(1+\sqrt{\rho}\frac{\sqrt{\nu}}{\sqrt{\nu}-R_{s}}\right)\frac{1}{\nu}\frac{d\nu}{dr}.\label{eq:photon-sphere-DE}
\end{equation}
This equation cannot be solved analytically. But we can obtain important
information on how to proceed from the plot of the potential and its
derivative that is presented in Fig. \ref{fig:Veff-Null-Exter}. From this figure it is evident that for the null case there seems
to be only one maximum, i.e., an unstable circular orbit, outside
of the black hole. This is similar to the classical case. We see that
the position of this photon sphere is very close to the classical
photon sphere, which is at $r_{\text{ph}}^{\text{class}}=3R_{s}/2$.
This observation lets us compute the approximate position of this
effective exterior photon sphere. To this end, we first expand $dV_{\text{eff}}/dr$
up to first order in $Q_{b}$ and $Q_{c}$. Then expand the resulting
expression for $r=3R_{s}/2+\delta r$ with $\delta r\to0$ up to the first order in $\delta r$. From here we solve this approximate
$dV_{\text{eff}}/dr=0$ for $\delta r$ and finally expand
this $\delta r$ up to the first order in $Q_{b}$ and $Q_{c}$. The
result is the approximate position of the photon sphere given by
\begin{equation}
r_{\text{ph}}^{\text{quat}}=\frac{3R_{s}}{2}-\frac{7Q_{b}}{9R_{s}}+\frac{64Q_{c}}{6561R_{s}^{5}}.
\end{equation}
As expected, the leading correction is only in $Q_{b}$, since this
is the quantum parameter that is important near the horizon.

\subsection{Timelike geodesics}

\begin{figure}
\begin{centering}
\includegraphics[scale=0.8]{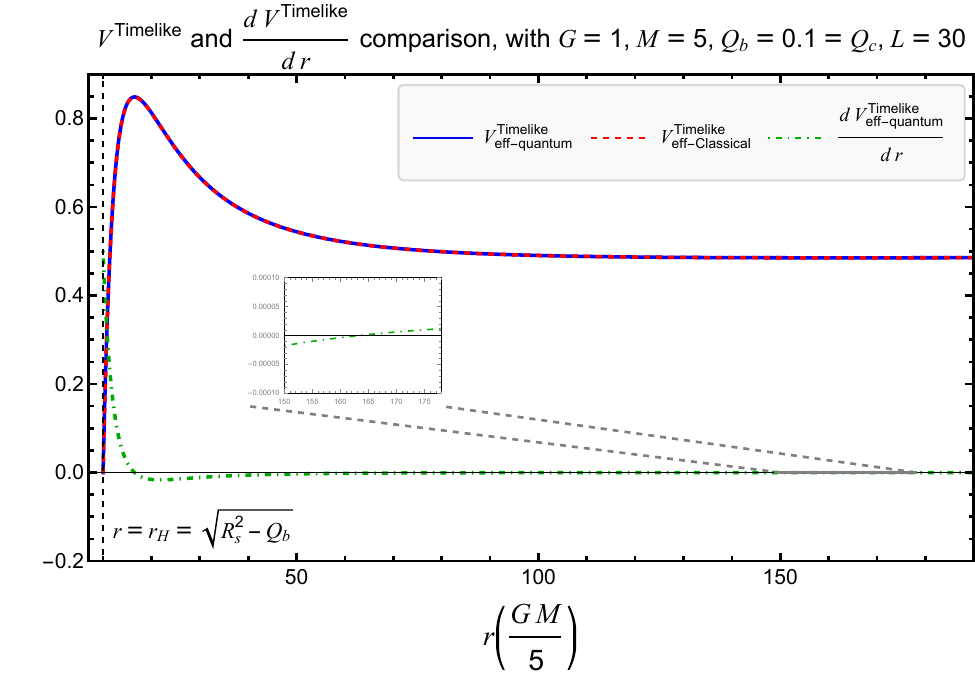}
\par\end{centering}
\caption{{\small{}The timelike effective potential for both the classical and
quantum cases, and the derivative of the effective potential in the
exterior. The quantum curve is qualitatively similar to the classical
one where there are two circular orbits in the exterior. \label{fig:Veff-Tlike-Exter}}}
\end{figure}

Setting $\varepsilon=1$ in (\ref{eq:circular-orbit-condition}) yields
the condition for circular orbits in the timelike case:
\begin{equation}
\left(g_{22}+L^{2}\right)g_{22}\frac{dg_{00}}{dr}-L^{2}g_{00}\frac{dg_{22}}{dr}=0.
\end{equation}
This equation is even more complicated than its null counterpart and
cannot be solve analytically. Nevertheless, we can again plot the effective
potential
\begin{equation}
V_{\text{eff}}=-\frac{1}{2}g_{00}\left[\frac{L^{2}}{g_{22}}+1\right]=\frac{1}{2}\sqrt{\frac{\nu}{\sqrt{\rho}}}\left(\sqrt{\nu}-R_{s}\right)\left[\frac{L^{2}}{\rho^{\frac{1}{4}}}+1\right]
\end{equation}
and its derivative. 
Figure~\ref{fig:Veff-Tlike-Exter}
shows the effective potential in the exterior of the black hole. In this region, seemingly there
are only two extrema for $V_{\text{eff}}$ corresponding to two circular
orbits, just as in the classical case. Although the location of these
in the quantum black hole are different from the classical ones by
a very small amount. This is
expected since all our computations until now show that the exterior
of this black hole exhibit similar qualitative behaviors as the classical
Schwarzschild black hole. 

\subsection{Painle\'{v}e-Gullstrand coordinates and infalling observers}

\subsubsection{Metric in the Painle\'{v}e-Gullstarnd coordinates}

The Painle\'{v}e-Gullstrand (PG) coordinates are derived from the
Schwarzschild coordinates by making a coordinate transformation $ t\to t_{\text{PG}}(r,t)$ where $t_{\text{PG}}$ is the proper time of the geodesics of an infalling observer with initial zero radial velocity and vanishing angular momentum. This condition means $U^{\mu}\partial_{\mu}t_{\text{PG}}=1$ with $U^{\theta}=0=U^{\phi}$. Setting the constants of motion $E=1$ and $L=0$, one obtains
\begin{equation}
dt_{\text{PG}}=dt+\sqrt{-g_{11}\left(1+\frac{1}{g_{00}}\right)}dr    
\end{equation}
After this transformation, the metric in the new coordinates is written as
\begin{equation}
ds^{2} = g_{00}^{(\text{PG})}dt_{\text{PG}}^{2}+2g_{01}^{(\text{PG})}dt_{\text{PG}}dr+g_{11}^{(\text{PG})}dr^{2}+g_{22}^{(\text{PG})}d\Omega^{2}\label{eq:PG-general}\, .
\end{equation}
For our metric, we obtain
\begin{align}
g_{00}^{(\text{PG})} & =g_{00}=-\frac{\sqrt{r^{2}+Q_{b}}}{\left(r^{8}+\frac{1}{4}Q_{c}R_{s}^{2}\right)^{\frac{1}{4}}}\left(\sqrt{r^{2}+Q_{b}}-R_{s}\right),\label{eq:g00-PG}\\
g_{01}^{(\text{PG})} & =-g_{00}\sqrt{-g_{11}\left(1+\frac{1}{g_{00}}\right)}=\sqrt{1+\frac{Q_{b}}{r^{2}}}\sqrt{1-\frac{\sqrt{r^{2}+Q_{b}}\left(\sqrt{r^{2}+Q_{b}}-R_{s}\right)}{\left(r^{8}+\frac{1}{4}Q_{c}R_{s}^{2}\right)^{\frac{1}{4}}}},\label{eq:g01-PG}\\
g_{11}^{(\text{PG})} & =-g_{00}g_{11}=1+\frac{Q_{b}}{r^{2}},\label{eq:g11-PG}\\
g_{22}^{(\text{PG})} & =g_{22}=\left(r^{8}+\frac{1}{4}Q_{c}R_{s}^{2}\right)^{\frac{1}{4}}.\label{eq:g22-PG}
\end{align}
This metric can be used to study the behavior of infalling observers
crossing the horizon. Notice that the inverse transformation from PG to diagonal in the interior matches precisely the metric \eqref{eq:met-int-r} with componets \eqref{eq:g00-imp-ext-1}-\eqref{eq:g22-imp-ext-1}, but with $t$ and $r$ switched.

\subsubsection{Velocity and proper time of the infalling geodesics}

\begin{figure}
\begin{centering}
\includegraphics[scale=0.8]{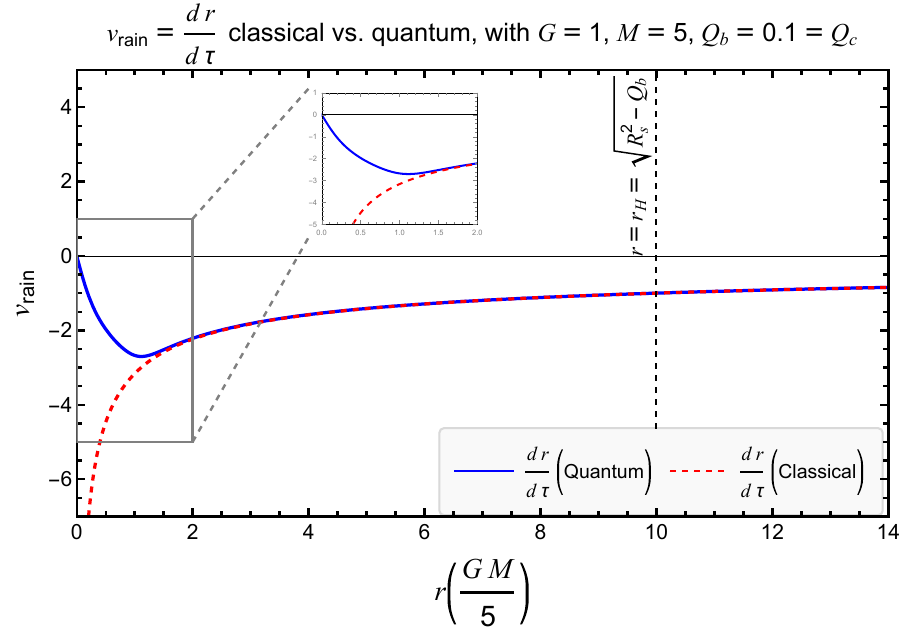}
\par\end{centering}
\caption{{\small{}Radial velocity of an infalling observer in Painle\'{v}e-Gullstrand coordinates for the classical and quantum cases. \label{fig:vRain}}}
\end{figure}

Using PG coordinates, we can compute certain interesting
aspects of an infalling observer. We first compute the velocity of
the observer falling into the black hole in its own frame (i.e., proper
time). For this, we replace the left-hand side of (\ref{eq:PG-general})
with $ds^{2}=-d\tau^{2}$, where $\tau$ is the proper time of the infalling
observer. We do the same for the right-hand side by replacing $dt_{\text{PG}}=d\tau$.
As a result, we obtain
\begin{equation}
-1=g_{00}^{(\text{PG})}+2g_{01}^{(\text{PG})}\frac{dr}{d\tau}+g_{11}^{(\text{PG})}\left(\frac{dr}{d\tau}\right)^{2},
\end{equation}
with the solution
\begin{equation}
v_{\text{rain}}=\frac{dr}{d\tau}=\frac{-g_{01}^{(\text{PG})}\pm\sqrt{-g_{00}^{(\text{PG})}g_{11}^{(\text{PG})}+\left(g_{01}^{(\text{PG})}\right)^{2}-g_{11}^{(\text{PG})}}}{g_{11}^{(\text{PG})}}.
\end{equation}
For our metric, both of the above solutions yield
\begin{equation}
v_{\text{rain}}=\frac{dr}{d\tau}=-\frac{r}{\sqrt{\nu}}\sqrt{1-\frac{\sqrt{\nu}\left(\sqrt{\nu}-R_{s}\right)}{\rho^{\frac{1}{4}}}}.\label{eq:drdtau-PG}
\end{equation}
In the classical limit $\rho_{\text{class}}=r^{8}$, and hence the
above velocity diverges at $r\to0$ as is well-known. However, quantum
corrections stop this expression from diverging, since $\rho$ does
not vanish in the quantum regime thanks to the presence of a quantum
term proportional to $Q_{c}$ in (\ref{eq:rho-def}). As can be seen
from Fig. \ref{fig:vRain}, $v_{\text{rain}}$ remains finite within
the black hole, and the derivative $dv_{\text{rain}}/dr$
changes sign close to $r=0$, as expected, to keep it from diverging
to $-\infty$. Furthermore, in the quantum case, the velocity at $r=0$
is $v_{\text{rain}}(r=0)=0$. From this, it is clear that the deciding factor in non-divergence
of the radial infalling velocity is $Q_{c}$ as expected since as
we mentioned before, $Q_{c}$ governs the important quantum effects
close to the $r=0$ region. 


Using (\ref{eq:drdtau-PG}), we also compute the
proper time for a typical radially infalling observer. If the observer
starts from an initial position $r_{i}$, the proper time it takes
for it to reach $r_{f}$ is computed using
\begin{equation}
\Delta\tau=\int_{r_{i}}^{r_{f}} \frac{1}{v_{\text{rain}}} dr.
\end{equation}
This integral is quite complicated and cannot be computed analytically. Numerical computations show that this proper time diverges whenever $r_f=0$. It is because the integrand diverges since, as discussed above, $v_{\text{rain}}(r=0)=0$.

One can see this analytically by expanding the integrand up to the first order in $Q_{b}$
and $Q_{c}$ and then evaluate the integral. The result is 
\begin{align}
\Delta\tau\approx & -\frac{2}{3}r\sqrt{\frac{r}{R_{s}}}+Q_{b}\sqrt{\frac{r}{R_{s}}}\left(\frac{1}{2r}-\frac{1}{R_{s}}\right)\nonumber \\
 & +\frac{Q_{c}}{r^{5}}\sqrt{\frac{R_{s}}{r}}\left(\frac{1}{208}\frac{R_{s}}{r}-\frac{1}{176}\right)\bigg|_{r_{i}}^{r_{f}}.
\end{align}
The classical
limit of this expression matches the classical Schwarzschild black
hole. In this limit, the falling observer will reach the singularity
located at $r_{f}=0$ in a finite proper time. However, in the quantum
regime, it takes infinite proper time to reach $r_{f}=0$ from any
initial point $r_{i}$.

\subsection{Expansion and Raychaudhuri equation}

\subsubsection{Expansion tensor and scalar}

To obtain the expansion scalar, we first need to find the so-called
expansion or deviation tensor. To find this tensor, we consider a congruence
of non-affinely parametrized null geodesics with null tangent vectors $k^{\mu}$ such that
\begin{align}
k_{\mu}k^{\mu}= & 0 \quad\textrm{and}\quad k^{\mu}\nabla_{\mu}k^{\nu}= \Omega\, k^\nu.\label{eq:k-mu-eqs}
\end{align}
Next, we decompose the spacetime
metric $g_{\mu\nu}$ into a longitudinal part and a part
$h_{\mu\nu}$ transverse to $k^{\mu}$.
For this, we need to introduce another auxiliary null vector field $\ell^{\mu}$ such that
\begin{align}
\ell_{\mu}\ell^{\mu}= 0 \quad\textrm{and}\quad k^{\mu}\ell_{\mu}=  -1.\label{eq:ll-kk-eqs}
\end{align}
In this way, we can decompose the spacetime metric $g_{\mu\nu}$ as
\begin{equation}
g_{\mu\nu}=h_{\mu\nu}-k_{\mu}\ell_{\nu}-k_{\nu}\ell_{\mu},
\end{equation}
where
\begin{align}
h_{\mu\nu}k^{\nu}=0 \quad\textrm{and}\quad h_{\mu\nu}\ell^{\nu}= 0.
\end{align}
The expansion tensor is then expressed as 
\begin{equation}
\Theta_{\mu\nu}=h^{\alpha}{}_{\mu}h^{\beta}{}_{v}\nabla_{\beta}k_{\alpha}.\label{eq:expan-tensor}
\end{equation}
This tensor can be decomposed into its irreducible parts 
\begin{equation}
\Theta_{\mu\nu}=\frac{1}{2}\theta h_{\mu\nu}+\sigma_{\mu\nu}+\omega_{\mu\nu},
\end{equation}
where the trace
\begin{equation}
\theta=g^{\mu\nu}\Theta_{\mu\nu},\label{eq:expan-scalar-gen}
\end{equation}
is the expansion scalar describing the expansion or compression  of the cross section of a congruence of null geodesics.
The traceless symmetric part 
\begin{equation}
\sigma_{\mu\nu}=\Theta_{(\mu\nu)}-\frac{1}{2}\theta h_{\mu\nu},\label{eq:shear-scalar-gen}
\end{equation}
is the shear describing how a circular cross section of the congruence
changes its shape, and the antisymmetric part
\begin{equation}
\omega_{\mu\nu}=\Theta_{[\mu\nu]},\label{eq:vorticity-scalar-gen}
\end{equation}
is the rotation or vorticity describing how the cross section rotates.

Starting with a null radial congruence described by 
\begin{equation}
k^{\mu}=\left(k^{0},k^{1},0,0\right),
\end{equation}
and the auxiliary field
\begin{equation}
\ell^{\mu}=\left(\ell^{0},\ell^{1},0,0\right),
\end{equation}
we can fix $k^0,\, k^1$ and $\ell^1$ by demanding
the left equation in \eqref{eq:k-mu-eqs} and the two equations in \eqref{eq:ll-kk-eqs}. In particular the condition $k^\mu k_\mu=0$ leads to two solutions for $k^1$, one corresponding to the outgoing and the other one to the ingoing radial null curves. The component $\ell^0>0$ remains arbitrary, and we choose it to be a constant, so that its choice only contributes to an overall scaling of expansion and Raychaudhuri equation. 

\begin{figure}
\begin{centering}
\includegraphics[scale=0.8]{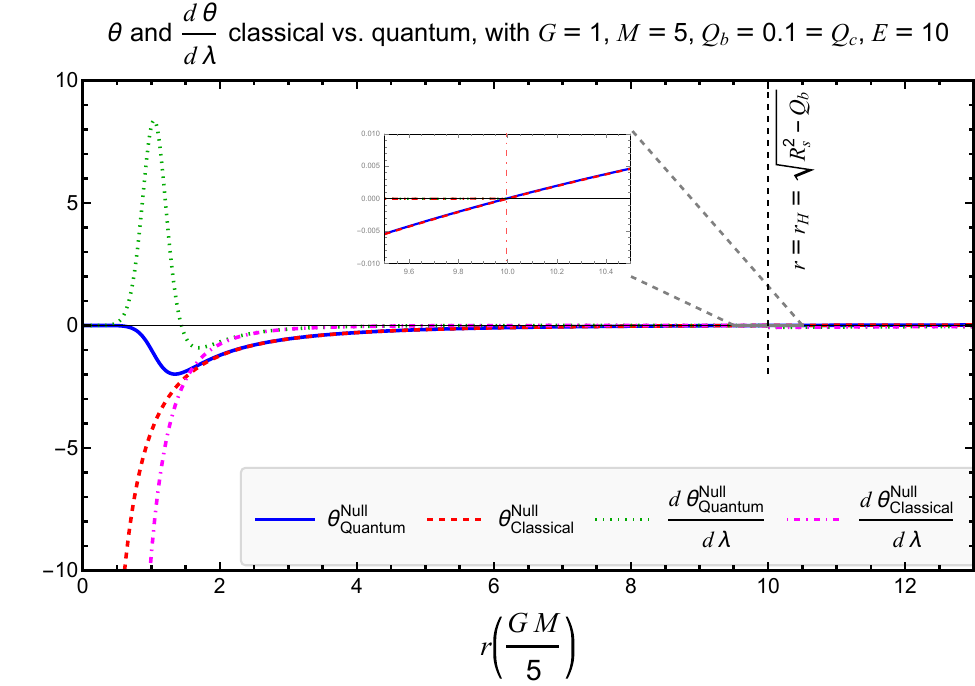}
\par\end{centering}
\caption{{\small{}The plot of the outgoing expansion scalar $\theta_+$ and the corresponding Raychaudhuri
equation $d\theta_+/d\lambda$ for a congruence of null radial geodesics. It is clear that
the quantum versions are finite everywhere particularly at $r=0$ where they vanish.
\label{fig:thetaRE}}}
\end{figure}

After this step, one can compute the expansion tensor in the PG coordinates using \eqref{eq:expan-tensor}. Using that, it is straightforward to compute the expansion, shear and vorticity using \eqref{eq:expan-scalar-gen}, \eqref{eq:shear-scalar-gen}, and \eqref{eq:vorticity-scalar-gen}, respectively as 
\begin{align}
\theta_{\pm} &= \frac{r^{8}}{\ell^{0}\sqrt{r^{2}+Q_{b}}\left(r^{8}+\frac{1}{4}Q_{c}R_{s}^{2}\right)}\left(\pm1-\sqrt{1+\frac{\sqrt{r^{2}+Q_{b}}\left(R_{s}-\sqrt{r^{2}+Q_{b}}\right)}{\left(r^{8}+\frac{1}{4}Q_{c}R_{s}^{2}\right)^{\frac{1}{4}}}}\right),\\
\sigma_{\mu\nu}\sigma^{\mu\nu} &= 0,\\
\omega_{\mu\nu}\omega^{\mu\nu} &= 0,
\end{align}
where $+$ and $-$ correspond to outgoing and ingoing curves, respectively. Notice that $\ell^0>0$ is an overall constant. As expected the classical limit, $Q_b\to 0$ and $Q_c\to 0$, of the expansion scalar matches that of the classical Schwarzschild black hole in PG coordinates, particularly for $\ell^0=1/2$,
\begin{equation}
\theta_{\pm}^{\text{class.}}=\frac{1}{\ell^{0}r}\left(\pm1-\sqrt{\frac{R_{s}}{r}}\right).
\end{equation}
We also
see that
\begin{equation}
\theta_{\pm}\left(r=0\right)=0.
\end{equation}
This, i.e., $\theta_\pm$ being finite everywhere, particularly at the location which used to be a singularity at $r=0$, is another strong indication of the resolution of singularity. Also note that at the horizon, we have
\begin{equation}
\theta_{\pm}(r=r_{H})=\frac{4(\pm1-1)\left(Q_{b}-R_{s}^{2}\right)^{4}}{\ell^{0}R_{s}\left[4\left(Q_{b}-R_{s}^{2}\right)^{4}+Q_{c}R_{s}^{2}\right]}\label{theta_pm_Horiz}
\end{equation}
which reduces to the classical expression for $Q_{b}\to0$ and $Q_{c}\to0$. Of particular importance is to notice that while $\theta_{-}$ is always negative (except at $r=0$), the outward expansion $\theta_{+}$ is positive in the exterior, becomes zero at the horizon $r=r_H$, and is negative in the interior. This shows that the black hole region is a trapped region as expected.

Using $k^1=dr/d\lambda$ above, where $\lambda$ is the non-affine parameter of the curves of the null geodesics, one can also integrate for the affine parameter along an infalling null radial geodesic as
\begin{equation}
\Delta\lambda=\int_{r_{i}}^{r_{f}} \frac{1}{k^1(r)} dr.
\end{equation}
This can be computed numerically and shows that it will diverge. Furthermore, a plot of $1/k^1$ itself shows that it diverges at $r=0$.

\subsubsection{Raychaudhuri equation}

The Raychaudhuri equation describes the evolution of the expansion
scalar in terms of the affine parameter of the geodesics. For a null
congruence we have
\begin{equation}
\frac{d\theta}{d\lambda}=-\frac{1}{2}\theta^{2}-\sigma_{\mu\nu}\sigma^{\mu\nu}+\omega_{\mu\nu}\omega^{\mu\nu}-R_{\mu\nu}k^{\mu}k^{\nu}.
\end{equation}
If we were considering the classical regime, clearly the last term
related to the null energy condition would have vanished due to
the fact that we are in vacuum. However, in the effective regime,
the equations of motion are not Einstein's equations and hence
we will not have $R_{\mu\nu}=0$ necessarily. In this regime, the Raychaudhuri
equation becomes
\begin{align}
\frac{d\theta_{\pm}}{d\lambda}= & \frac{r^{8}\left(r^{10}-\frac{7}{4}Q_{c}R_{s}^{2}r^{2}-2Q_{c}Q_{b}R_{s}^{2}\right)}{\left(r^{2}+Q_{b}\right)^{2}\left(r^{8}+\frac{1}{4}Q_{c}R_{s}^{2}\right)^{2}\left(\ell^{0}\right)^{2}}\times\nonumber \\
 & \left[\pm\sqrt{1+\frac{\sqrt{r^{2}+Q_{b}}\left(R_{s}-\sqrt{r^{2}+Q_{b}}\right)}{\left(r^{8}+\frac{1}{4}Q_{c}R_{s}^{2}\right)^{1/4}}}-\frac{\sqrt{r^{2}+Q_{b}}\left(R_{s}-\sqrt{r^{2}+Q_{b}}\right)}{2\left(r^{8}+\frac{1}{4}Q_{c}R_{s}^{2}\right)^{1/4}}-1\right]
\end{align}
This expression never diverges and in particular becomes zero at $r=0$. Furthermore, the classical limit of this expression yields
\begin{equation}
\frac{d\theta_{\pm}}{d\lambda}=\frac{1}{2r^{3}\left(\ell^{0}\right)^{2}}\left[\pm 2 \, \sqrt{rR_{s}}-r-R_{s}\right]
\end{equation}
which clearly diverges at $r\to 0$.


A plot of both $\theta_+$ and $d\theta_+/d\lambda$ for outgoing rays is
presented in Fig. \ref{fig:thetaRE}. The regularity of both $\theta$
and $d\theta/d\lambda$ was also observed before in
the models that only considered the interior \cite{Blanchette:2021vid,Bosso:2020ztk,Rastgoo:2022mks,Bosso:2023fnb}.

The finiteness, and in fact vanishing, of both null $\theta_{\pm}$
and $d\theta_{\pm}/d\lambda$ over the entire spacetime, together with regularity of the Kretschmann
scalar everywhere in this spacetime indicates that the singularity
is resolved in this quantum black hole model.

\section{Discussion and conclusions \label{sec:Conclusion}}

In this work, we have examined whether one can analytically continue
the interior metric of a GUP-inspired black hole introduced
earlier~\cite{Bosso:2020ztk} to the full spacetime, via switching the radial
spacelike and the timelike coordinates $t\leftrightarrow r$. This
would be similar to the classical Schwarzschild case where one can
simply obtain the interior from the exterior, and vice versa, in this
way. Our analysis showed that the resulting extended metric and consequently
the Kretschmann scalar have serious issues in the asymptotic $r\to\infty$
region: $g_{00}$ will not be in the desired form and the Kretschmann
scalar will fall of as $r^{-4}$ instead of $r^{-6}$.
This issue is similar to the case of \cite{Ashtekar:2020ckv,Bouhmadi-Lopez:2019hpp}. 

In order to remedy these issues, we have introduced an improved scheme, borrowed from the techniques
used in loop quantum gravity, whereby one makes the previously constant
quantum parameters of the model momentum-dependent. In this specific
model, the minimal uncertainty parameter are $\beta_{b}$ and $\beta_{c}$,
and we modify them to $\bar{\beta}_{b}=\beta_{b}L_0^4/p_{b}^{2}$
and $\bar{\beta}_{c}=\beta_{c}L_0^4/p_{c}^{2}$, where $p_{b}$
and $p_{c}$ are momenta of the theory which correspond to the components
of the densitized triads; essentially the components of the metric.
Remarkably, this rather simple prescription, not only results in all
the desired asymptotic behaviors, but also renders the black hole regular.
Furthermore, the Kretschmann scalar vanishes at $r\to\infty$ as in the classicla case, and
more importantly its expansion up to zero'th order in quantum
parameters exactly matches that of the Schwarzschild metric. In addition,
all the classical limits ($\beta_{b}\to0$ and $\beta_{c}\to0$) match the
classical Schwarzschild spacetime.

We find that in order for the metric components to always remain real, $\beta_{b}$ and $\beta_{c}$ should
be negative. This is the same result that was previously obtained if one wants to resolve the singularity~\cite{Bosso:2020ztk}. In this work, this
condition is derived more systematically and is much stronger (demanding
metric reality and not finiteness of the Kretschmann scalar). The
condition then leads directly to the resolution of the singularity
by rendering the denominator of the Kretschmann scalar positive and
nonzero for any value of the radial coordinate $r$, including the horizon.

In Schwarzschild coordinates in which the metric function $g_{22} = \bar{r}^2$ for some $\bar{r}$ coordinate,
we find a minimum mass-dependent size of 2-spheres given by quantum parameter $Q_c$, while the other quantum parameter $Q_b$ is responsible for quantum effects near the horizon.

Finally, we computed the null expansion and the Raychaudhuri
equation for a congruence of null geodesics and show that they both
are regular everywhere in this spacetime, which together
with regularity of the Ketschmann scalar, indicates that the singularity
is resolved in this quantum black hole spacetime.

\acknowledgments{
The authors acknowledge the support of the Natural Sciences and Engineering
Research Council of Canada (NSERC).
Nous remercions le Conseil de recherches en sciences naturelles et en
g{\'e}nie du Canada (CRSNG) de son soutien.
}


\appendix


\section{Equations of motion of the interior\label{app:Interior-EoM}}
Given the classical equations of motion ($i=b,\,c$)
\begin{align}
\dot{q}_{i}= & \left\{ q_{i},H\right\} _{\text{cl.}}, & \dot{p}_{i}= & \left\{ p_{i},H\right\} _{\text{cl.}},
\end{align}
using the classical Poisson brackets \eqref{eq:PB-classic}, if the
classical algebra change to the modified algebra as 
\begin{equation}
\left\{ q_{i},p_{j}\right\} _{\text{eff.}}=\left\{ q_{i},p_{j}\right\} _{\text{cl.}}\left(1+F\left(q,p,\beta_{i}\right)\right)\delta_{ij},
\end{equation}
the new equations of motion become 
\begin{align}
\dot{q}_{i}= & \left\{ q_{i},H\right\} _{\text{eff.}}=\left\{ q_{i},H\right\} _{\text{cl.}}\left(1+F\left(q,p,\beta_{i}\right)\right), & \dot{p}_{i}= & \left\{ p_{i},H\right\} _{\text{eff.}}=\left\{ p_{i},H\right\} _{\text{cl.}}\left(1+F\left(q,p,\beta_{i}\right)\right).
\end{align}
For the non-improved interior case where the solutions are Eqs. \eqref{eq:GUP-int-b}-\eqref{eq:GUP-int-pc},
these EoM are
\begin{align}
\dot{\tilde{b}}= & -\frac{\tilde{b}^{2}+\gamma^{2}}{2\tilde{b}}\left(1+\beta_{b}\tilde{b}^{2}\right),\\
\dot{\tilde{p}}_{b}=~ & \frac{\tilde{p}_{b}\left(\tilde{b}^{2}-\gamma^{2}\right)}{2\tilde{b}^{2}}\left(1+\beta_{b}\tilde{b}^{2}\right),\\
\dot{\tilde{c}}= & -2\tilde{c}\left(1+\beta_{c}\tilde{c}^{2}\right),\\
\dot{\tilde{p}}_{c}=~ & 2\tilde{p}_{c}\left(1+\beta_{c}\tilde{c}^{2}\right).
\end{align}
For the improved interior whose solutions are presented in \eqref{eq:b-int-extended}-\eqref{eq:pc-int-extended},
the corresponding EoM are
\begin{align}
\dot{b}= & -\frac{b^{2}+\gamma^{2}}{2b}\left(1+\frac{\beta_{b}L_{0}^{4}}{p_{b}^{2}}b^{2}\right),\\
\dot{p}_{b}=~ & \frac{p_{b}\left(b^{2}-\gamma^{2}\right)}{2b^{2}}\left(1+\frac{\beta_{b}L_{0}^{4}}{p_{b}^{2}}b^{2}\right),\\
\dot{c}= & -2c\left(1+\frac{\beta_{c}L_{0}^{4}}{p_{c}^{2}}c^{2}\right),\\
\dot{p}_{c}=~ & 2p_{c}\left(1+\frac{\beta_{c}L_{0}^{4}}{p_{c}^{2}}c^{2}\right).
\end{align}

\bibliographystyle{JHEP}
\bibliography{mainbib}

\providecommand{\href}[2]{#2}\begingroup\raggedright\begin{thebibliography}{10}

\bibitem{Addazi:2021xuf}
A.~Addazi et~al., \emph{{Quantum gravity phenomenology at the dawn of the
  multi-messenger era\textemdash{}A review}},
  \href{https://doi.org/10.1016/j.ppnp.2022.103948}{\emph{Prog. Part. Nucl.
  Phys.} {\bfseries 125} (2022) 103948}
  [\href{https://arxiv.org/abs/2111.05659}{{\ttfamily 2111.05659}}].

\bibitem{LISA:2022kgy}
{\scshape LISA} collaboration, \emph{{New horizons for fundamental physics with
  LISA}}, \href{https://doi.org/10.1007/s41114-022-00036-9}{\emph{Living Rev.
  Rel.} {\bfseries 25} (2022) 4}
  [\href{https://arxiv.org/abs/2205.01597}{{\ttfamily 2205.01597}}].

\bibitem{LISACosmologyWorkingGroup:2022jok}
{\scshape LISA Cosmology Working Group} collaboration, \emph{{Cosmology with
  the Laser Interferometer Space Antenna}},
  \href{https://doi.org/10.1007/s41114-023-00045-2}{\emph{Living Rev. Rel.}
  {\bfseries 26} (2023) 5} [\href{https://arxiv.org/abs/2204.05434}{{\ttfamily
  2204.05434}}].

\bibitem{AlvesBatista:2023wqm}
R.~Alves~Batista et~al., \emph{{White Paper and Roadmap for Quantum Gravity
  Phenomenology in the Multi-Messenger Era}},
  \href{https://arxiv.org/abs/2312.00409}{{\ttfamily 2312.00409}}.

\bibitem{Thiemann:2007pyv}
T.~Thiemann, \emph{{Modern Canonical Quantum General Relativity}}, Cambridge
  Monographs on Mathematical Physics. Cambridge University Press, 2007,
  \href{https://doi.org/10.1017/CBO9780511755682}{10.1017/CBO9780511755682}.

\bibitem{Ashtekar:2005qt}
A.~Ashtekar and M.~Bojowald, \emph{{Quantum geometry and the Schwarzschild
  singularity}}, \href{https://doi.org/10.1088/0264-9381/23/2/008}{\emph{Class.
  Quant. Grav.} {\bfseries 23} (2006) 391}
  [\href{https://arxiv.org/abs/gr-qc/0509075}{{\ttfamily gr-qc/0509075}}].

\bibitem{Bohmer:2007wi}
{B\"ohmer, Christian G. and Vandersloot, Kevin}, \emph{{Loop Quantum Dynamics
  of the Schwarzschild Interior}},
  \href{https://doi.org/10.1103/PhysRevD.76.104030}{\emph{Phys. Rev. D}
  {\bfseries 76} (2007) 104030}
  [\href{https://arxiv.org/abs/0709.2129}{{\ttfamily 0709.2129}}].

\bibitem{Chiou:2008nm}
D.-W. Chiou, \emph{{Phenomenological loop quantum geometry of the Schwarzschild
  black hole}}, \href{https://doi.org/10.1103/PhysRevD.78.064040}{\emph{Phys.
  Rev. D} {\bfseries 78} (2008) 064040}
  [\href{https://arxiv.org/abs/0807.0665}{{\ttfamily 0807.0665}}].

\bibitem{Morales-Tecotl:2018ugi}
H.~A. Morales-T\'ecotl, S.~Rastgoo and J.~C. Ruelas, \emph{{Effective dynamics
  of the Schwarzschild black hole interior with inverse triad corrections}},
  \href{https://doi.org/10.1016/j.aop.2021.168401}{\emph{Annals Phys.}
  {\bfseries 426} (2021) 168401}
  [\href{https://arxiv.org/abs/1806.05795}{{\ttfamily 1806.05795}}].

\bibitem{Blanchette:2020kkk}
K.~Blanchette, S.~Das, S.~Hergott and S.~Rastgoo, \emph{{Black hole singularity
  resolution via the modified Raychaudhuri equation in loop quantum gravity}},
  \href{https://doi.org/10.1103/PhysRevD.103.084038}{\emph{Phys. Rev. D}
  {\bfseries 103} (2021) 084038}
  [\href{https://arxiv.org/abs/2011.11815}{{\ttfamily 2011.11815}}].

\bibitem{Kelly:2020uwj}
J.~G. Kelly, R.~Santacruz and E.~Wilson-Ewing, \emph{{Effective loop quantum
  gravity framework for vacuum spherically symmetric space-times}},
  \href{https://arxiv.org/abs/2006.09302}{{\ttfamily 2006.09302}}.

\bibitem{Gambini:2020nsf}
R.~Gambini, J.~Olmedo and J.~Pullin, \emph{{Spherically symmetric loop quantum
  gravity: analysis of improved dynamics}},
  \href{https://doi.org/10.1088/1361-6382/aba842}{\emph{Class. Quant. Grav.}
  {\bfseries 37} (2020) 205012}
  [\href{https://arxiv.org/abs/2006.01513}{{\ttfamily 2006.01513}}].

\bibitem{Ashtekar:2018lag}
A.~Ashtekar, J.~Olmedo and P.~Singh, \emph{{Quantum Transfiguration of Kruskal
  Black Holes}},
  \href{https://doi.org/10.1103/PhysRevLett.121.241301}{\emph{Phys. Rev. Lett.}
  {\bfseries 121} (2018) 241301}
  [\href{https://arxiv.org/abs/1806.00648}{{\ttfamily 1806.00648}}].

\bibitem{Bodendorfer:2019nvy}
N.~Bodendorfer, F.~M. Mele and J.~M\"{u}nch, \emph{{(b,v)-Type Variables for
  Black to White Hole Transitions in Effective Loop Quantum Gravity}},
  \href{https://arxiv.org/abs/1911.12646}{{\ttfamily 1911.12646}}.

\bibitem{Modesto:2008im}
L.~Modesto, \emph{{Semiclassical loop quantum black hole}},
  \href{https://doi.org/10.1007/s10773-010-0346-x}{\emph{Int. J. Theor. Phys.}
  {\bfseries 49} (2010) 1649}
  [\href{https://arxiv.org/abs/0811.2196}{{\ttfamily 0811.2196}}].

\bibitem{Bojowald:2020dkb}
M.~Bojowald, \emph{{Black-Hole Models in Loop Quantum Gravity}},
  \href{https://doi.org/10.3390/universe6080125}{\emph{Universe} {\bfseries 6}
  (2020) 125} [\href{https://arxiv.org/abs/2009.13565}{{\ttfamily
  2009.13565}}].

\bibitem{Gambini:2009vp}
R.~Gambini, J.~Pullin and S.~Rastgoo, \emph{New variables for 1+1 dimensional
  gravity},
  \href{https://doi.org/10.1088/0264-9381/27/2/025002}{\emph{Class.Quant.Grav.}
  {\bfseries 27} (2010) 025002}
  [\href{https://arxiv.org/abs/0909.0459}{{\ttfamily 0909.0459}}].

\bibitem{Corichi:2015vsa}
A.~Corichi, A.~Karami, S.~Rastgoo and T.~Vuka\v{s}inac, \emph{{Constraint Lie
  algebra and local physical Hamiltonian for a generic 2D dilatonic model}},
  \href{https://doi.org/10.1088/0264-9381/33/3/035011}{\emph{Class. Quant.
  Grav.} {\bfseries 33} (2016) 035011}
  [\href{https://arxiv.org/abs/1508.03036}{{\ttfamily 1508.03036}}].

\bibitem{Corichi:2016nkp}
A.~Corichi, J.~Olmedo and S.~Rastgoo,
  \emph{Callan-{G}iddings-{H}arvey-{S}trominger vacuum in loop quantum gravity
  and singularity resolution},
  \href{https://doi.org/10.1103/PhysRevD.94.084050}{\emph{Phys.Rev.D}
  {\bfseries 94} (2016) 084050}
  [\href{https://arxiv.org/abs/1608.06246}{{\ttfamily 1608.06246}}].

\bibitem{Ashtekar:2002sn}
A.~Ashtekar, S.~Fairhurst and J.~L. Willis, \emph{{Quantum gravity, shadow
  states, and quantum mechanics}},
  \href{https://doi.org/10.1088/0264-9381/20/6/302}{\emph{Class. Quant. Grav.}
  {\bfseries 20} (2003) 1031}
  [\href{https://arxiv.org/abs/gr-qc/0207106}{{\ttfamily gr-qc/0207106}}].

\bibitem{Morales-Tecotl:2016ijb}
H.~A. Morales-T\'{e}cotl, S.~Rastgoo and J.~C. Ruelas, \emph{Path integral
  polymer propagator of relativistic and nonrelativistic particles},
  \href{https://doi.org/10.1103/PhysRevD.95.065026}{\emph{Phys.Rev.D}
  {\bfseries 95} (2017) 065026}
  [\href{https://arxiv.org/abs/1608.04498}{{\ttfamily 1608.04498}}].

\bibitem{Tecotl:2015cya}
H.~A. Morales-T\'{e}cotl, D.~H. Orozco-Borunda and S.~Rastgoo, \emph{Polymer
  quantization and the saddle point approximation of partition functions},
  \href{https://doi.org/10.1103/PhysRevD.92.104029}{\emph{Phys.Rev.D}
  {\bfseries 92} (2015) 104029}
  [\href{https://arxiv.org/abs/1507.08651}{{\ttfamily 1507.08651}}].

\bibitem{Kempf:1994su}
A.~Kempf, G.~Mangano and R.~B. Mann, \emph{{Hilbert space representation of the
  minimal length uncertainty relation}},
  \href{https://doi.org/10.1103/PhysRevD.52.1108}{\emph{Phys. Rev. D}
  {\bfseries 52} (1995) 1108}
  [\href{https://arxiv.org/abs/hep-th/9412167}{{\ttfamily hep-th/9412167}}].

\bibitem{Bosso:2023aht}
P.~Bosso, G.~G. Luciano, L.~Petruzziello and F.~Wagner, \emph{{30 years in: Quo
  vadis generalized uncertainty principle?}},
  \href{https://doi.org/10.1088/1361-6382/acf021}{\emph{Class. Quant. Grav.}
  {\bfseries 40} (2023) 195014}
  [\href{https://arxiv.org/abs/2305.16193}{{\ttfamily 2305.16193}}].

\bibitem{Bosso:2020ztk}
P.~Bosso, O.~Obreg\'on, S.~Rastgoo and W.~Yupanqui, \emph{{Deformed algebra and
  the effective dynamics of the interior of black holes}},
  \href{https://doi.org/10.1088/1361-6382/ac025f}{\emph{Class. Quant. Grav.}
  {\bfseries 38} (2021) 145006}
  [\href{https://arxiv.org/abs/2012.04795}{{\ttfamily 2012.04795}}].

\bibitem{Blanchette:2021vid}
K.~Blanchette, S.~Das and S.~Rastgoo, \emph{{Effective GUP-modified
  Raychaudhuri equation and black hole singularity: four models}},
  \href{https://doi.org/10.1007/JHEP09(2021)062}{\emph{JHEP} {\bfseries 09}
  (2021) 062} [\href{https://arxiv.org/abs/2105.11511}{{\ttfamily
  2105.11511}}].

\bibitem{Rastgoo:2022mks}
S.~Rastgoo and S.~Das, \emph{{Probing the Interior of the Schwarzschild Black
  Hole Using Congruences: LQG vs. GUP}},
  \href{https://doi.org/10.3390/universe8070349}{\emph{Universe} {\bfseries 8}
  (2022) 349} [\href{https://arxiv.org/abs/2205.03799}{{\ttfamily
  2205.03799}}].

\bibitem{Bosso:2023fnb}
P.~Bosso, O.~Obreg\'on, S.~Rastgoo and W.~Yupanqui, \emph{{Black hole interior
  quantization: a minimal uncertainty approach}},
  \href{https://arxiv.org/abs/2310.04600}{{\ttfamily 2310.04600}}.

\bibitem{Lambiase:2017adh}
G.~Lambiase and F.~Scardigli, \emph{{Lorentz violation and generalized
  uncertainty principle}},
  \href{https://doi.org/10.1103/PhysRevD.97.075003}{\emph{Phys. Rev. D}
  {\bfseries 97} (2018) 075003}
  [\href{https://arxiv.org/abs/1709.00637}{{\ttfamily 1709.00637}}].

\bibitem{Scardigli:2014qka}
F.~Scardigli and R.~Casadio, \emph{{Gravitational tests of the Generalized
  Uncertainty Principle}},
  \href{https://doi.org/10.1140/epjc/s10052-015-3635-y}{\emph{Eur. Phys. J. C}
  {\bfseries 75} (2015) 425} [\href{https://arxiv.org/abs/1407.0113}{{\ttfamily
  1407.0113}}].

\bibitem{Casadio:2016fev}
R.~Casadio, A.~Giugno and A.~Giusti, \emph{{Global and Local Horizon Quantum
  Mechanics}}, \href{https://doi.org/10.1007/s10714-017-2198-7}{\emph{Gen. Rel.
  Grav.} {\bfseries 49} (2017) 32}
  [\href{https://arxiv.org/abs/1605.06617}{{\ttfamily 1605.06617}}].

\bibitem{Jizba:2009qf}
P.~Jizba, H.~Kleinert and F.~Scardigli, \emph{{Uncertainty Relation on World
  Crystal and its Applications to Micro Black Holes}},
  \href{https://doi.org/10.1103/PhysRevD.81.084030}{\emph{Phys. Rev. D}
  {\bfseries 81} (2010) 084030}
  [\href{https://arxiv.org/abs/0912.2253}{{\ttfamily 0912.2253}}].

\bibitem{Carr:2015nqa}
B.~J. Carr, J.~Mureika and P.~Nicolini, \emph{{Sub-Planckian black holes and
  the Generalized Uncertainty Principle}},
  \href{https://doi.org/10.1007/JHEP07(2015)052}{\emph{JHEP} {\bfseries 07}
  (2015) 052} [\href{https://arxiv.org/abs/1504.07637}{{\ttfamily
  1504.07637}}].

\bibitem{Ong:2018zqn}
Y.~C. Ong, \emph{{Generalized Uncertainty Principle, Black Holes, and White
  Dwarfs: A Tale of Two Infinities}},
  \href{https://doi.org/10.1088/1475-7516/2018/09/015}{\emph{JCAP} {\bfseries
  09} (2018) 015} [\href{https://arxiv.org/abs/1804.05176}{{\ttfamily
  1804.05176}}].

\bibitem{Buoninfante:2019fwr}
L.~Buoninfante, G.~G. Luciano and L.~Petruzziello, \emph{{Generalized
  Uncertainty Principle and Corpuscular Gravity}},
  \href{https://doi.org/10.1140/epjc/s10052-019-7164-y}{\emph{Eur. Phys. J. C}
  {\bfseries 79} (2019) 663}
  [\href{https://arxiv.org/abs/1903.01382}{{\ttfamily 1903.01382}}].

\bibitem{Modesto:2005zm}
L.~Modesto, \emph{{Loop quantum black hole}},
  \href{https://doi.org/10.1088/0264-9381/23/18/006}{\emph{Class. Quant. Grav.}
  {\bfseries 23} (2006) 5587}
  [\href{https://arxiv.org/abs/gr-qc/0509078}{{\ttfamily gr-qc/0509078}}].

\bibitem{Ashtekar:2020ckv}
A.~Ashtekar and J.~Olmedo, \emph{{Properties of a recent quantum extension of
  the Kruskal geometry}},
  \href{https://doi.org/10.1142/S0218271820500765}{\emph{Int. J. Mod. Phys. D}
  {\bfseries 29} (2020) 2050076}
  [\href{https://arxiv.org/abs/2005.02309}{{\ttfamily 2005.02309}}].

\bibitem{Bouhmadi-Lopez:2019hpp}
M.~Bouhmadi-L\'opez, S.~Brahma, C.-Y. Chen, P.~Chen and D.-h. Yeom,
  \emph{{Asymptotic non-flatness of an effective black hole model based on loop
  quantum gravity}},
  \href{https://doi.org/10.1016/j.dark.2020.100701}{\emph{Phys. Dark Univ.}
  {\bfseries 30} (2020) 100701}
  [\href{https://arxiv.org/abs/1902.07874}{{\ttfamily 1902.07874}}].

\bibitem{Rashidi:2015rro}
R.~Rashidi, \emph{{Generalized uncertainty principle and the maximum mass of
  ideal white dwarfs}},
  \href{https://doi.org/10.1016/j.aop.2016.09.005}{\emph{Annals Phys.}
  {\bfseries 374} (2016) 434}
  [\href{https://arxiv.org/abs/1512.06356}{{\ttfamily 1512.06356}}].

\bibitem{Mathew:2020wnx}
A.~Mathew and M.~K. Nandy, \emph{{Existence of Chandrasekhar's limit in GUP
  white dwarfs}},  \href{https://arxiv.org/abs/2002.08360}{{\ttfamily
  2002.08360}}.

\bibitem{Ashtekar:2006wn}
A.~Ashtekar, T.~Pawlowski and P.~Singh, \emph{{Quantum Nature of the Big Bang:
  Improved dynamics}},
  \href{https://doi.org/10.1103/PhysRevD.74.084003}{\emph{Phys. Rev.}
  {\bfseries D74} (2006) 084003}
  [\href{https://arxiv.org/abs/gr-qc/0607039}{{\ttfamily gr-qc/0607039}}].

\bibitem{Chiou:2012pg}
D.-W. Chiou, W.-T. Ni and A.~Tang, \emph{{Loop quantization of spherically
  symmetric midisuperspaces and loop quantum geometry of the maximally extended
  Schwarzschild spacetime}},  \href{https://arxiv.org/abs/1212.1265}{{\ttfamily
  1212.1265}}.

\bibitem{Simpson:2018tsi}
A.~Simpson and M.~Visser, \emph{{Black-bounce to traversable wormhole}},
  \href{https://doi.org/10.1088/1475-7516/2019/02/042}{\emph{JCAP} {\bfseries
  02} (2019) 042} [\href{https://arxiv.org/abs/1812.07114}{{\ttfamily
  1812.07114}}].

\bibitem{PhysRev.129.1873}
A.~Komar, \emph{Positive-definite energy density and global consequences for
  general relativity},
  \href{https://doi.org/10.1103/PhysRev.129.1873}{\emph{Phys. Rev.} {\bfseries
  129} (1963) 1873}.

\bibitem{Carroll:book}
S.~M. {C}arroll, \emph{{Spacetime And Geometry An Introduction To General
  Relativity}}. Addison Wesley, 2004.

\bibitem{PhysRev.116.1322}
R.~Arnowitt, S.~Deser and C.~W. Misner, \emph{Dynamical structure and
  definition of energy in general relativity},
  \href{https://doi.org/10.1103/PhysRev.116.1322}{\emph{Phys. Rev.} {\bfseries
  116} (1959) 1322}.

\bibitem{Gourgoulhon:2007ue}
E.~Gourgoulhon, \emph{{3+1 formalism and bases of numerical relativity}},
  \href{https://arxiv.org/abs/gr-qc/0703035}{{\ttfamily gr-qc/0703035}}.

\bibitem{PhysRev.136.B571}
C.~W. Misner and D.~H. Sharp, \emph{Relativistic equations for adiabatic,
  spherically symmetric gravitational collapse},
  \href{https://doi.org/10.1103/PhysRev.136.B571}{\emph{Phys. Rev.} {\bfseries
  136} (1964) B571}.

\bibitem{Faraoni:2020mdf}
V.~Faraoni, A.~Giusti and T.~F. Bean, \emph{{Asymptotic flatness and Hawking
  quasilocal mass}},
  \href{https://doi.org/10.1103/PhysRevD.103.044026}{\emph{Phys. Rev. D}
  {\bfseries 103} (2021) 044026}
  [\href{https://arxiv.org/abs/2010.00069}{{\ttfamily 2010.00069}}].

\end{thebibliography}\endgroup

\end{document}